\documentclass[sigconf]{acmart}

\AtBeginDocument{%
  }

\renewcommand\footnotetextcopyrightpermission[1]{} 
\usepackage{tikz}
\usepackage{amsmath,amssymb,amsfonts}
\usepackage{subfigure}
\usepackage{algorithm}
\usepackage{algpseudocode}
\usepackage{multirow, array}
\usepackage{adjustbox}
\usepackage{graphicx}
\usepackage{pifont}
\usepackage{float}
\usepackage{enumitem}

\DeclareMathOperator{\argmax}{argmax} 
\usepackage{filecontents}

\usepackage{xcolor}
\newcommand{\circled}[1]{\tikz[baseline=(char.base)]{\node[shape=circle,fill=black!100,text=white,inner sep=-0.pt, font=\fontsize{7}{0}\selectfont] (char) {#1};}}

\usepackage{draftwatermark}
\SetWatermarkText{Preview} 
\SetWatermarkLightness{0.9} 
\SetWatermarkScale{1.0} 

\copyrightyear{2024} 
\acmYear{2024} 
\setcopyright{acmcopyright}
\acmConference[SIGMOD'24] {Proceedings of In Proceedings of the 2024 International Conference on Management of Data}{June 9--15, 2024}{Santiago, Chile.}
\acmBooktitle{Proceedings of In Proceedings of the 2024 International Conference on Management of Data (SIGMOD '24), June 9--15, 2024, Santiago, Chile}
\acmPrice{15.00}
\acmISBN{978-1-4503-XXXX-X} 
\acmDOI{https://doi.org/XXXXXXX.XXXXXXX}
\begin{document}

\title{DGC: Training Dynamic Graphs with Spatio-Temporal Non-Uniformity using Graph Partitioning by Chunks}

\author{Fahao Chen, Peng Li}
\authornote{Peng Li is the corresponding author.}
\affiliation{%
  \institution{School of Computer Science and Engineering\\The University of Aizu}
\country{Japan}
}
\email{{d8232101, pengli}@u-aizu.ac.jp}


\author{Celimuge Wu}
\affiliation{%
  \institution{Department of Computer and Network Engineering\\ Graduate School of Informatics and Engineering\\University of Electro-Communications}
\country{Japan}
}
\email{celimuge@uec.ac.jp}

\begin{abstract}
Dynamic Graph Neural Network (DGNN) has shown a strong capability of learning dynamic graphs by exploiting both spatial and temporal features. Although DGNN has recently received considerable attention by AI community and various DGNN models have been proposed, building a distributed system for efficient DGNN training is still challenging. It has been well recognized that how to partition the dynamic graph and assign workloads to multiple GPUs plays a critical role in training acceleration. Existing works partition a dynamic graph into snapshots or temporal sequences, which only work well when the graph has uniform spatio-temporal structures. However, dynamic graphs in practice are not uniformly structured, with some snapshots being very dense while others are sparse. To address this issue, we propose DGC, a distributed DGNN training system that achieves a 1.25$\times$ - 7.52$\times$ speedup over the state-of-the-art in our testbed. DGC's success stems from a new graph partitioning method that partitions dynamic graphs into chunks, which are essentially subgraphs with modest training workloads and few inter connections. This partitioning algorithm is based on graph coarsening, which can run very fast on large graphs. In addition, DGC has a highly efficient run-time, powered by the proposed chunk fusion and adaptive stale aggregation techniques. Extensive experimental results on 3 typical DGNN models and 4 popular dynamic graph datasets are presented to show the effectiveness of DGC.
\end{abstract}





\maketitle

\section{Introduction} \label{introduction}

\begin{figure}[t] 
\begin{center}
\includegraphics[width=0.4\textwidth]{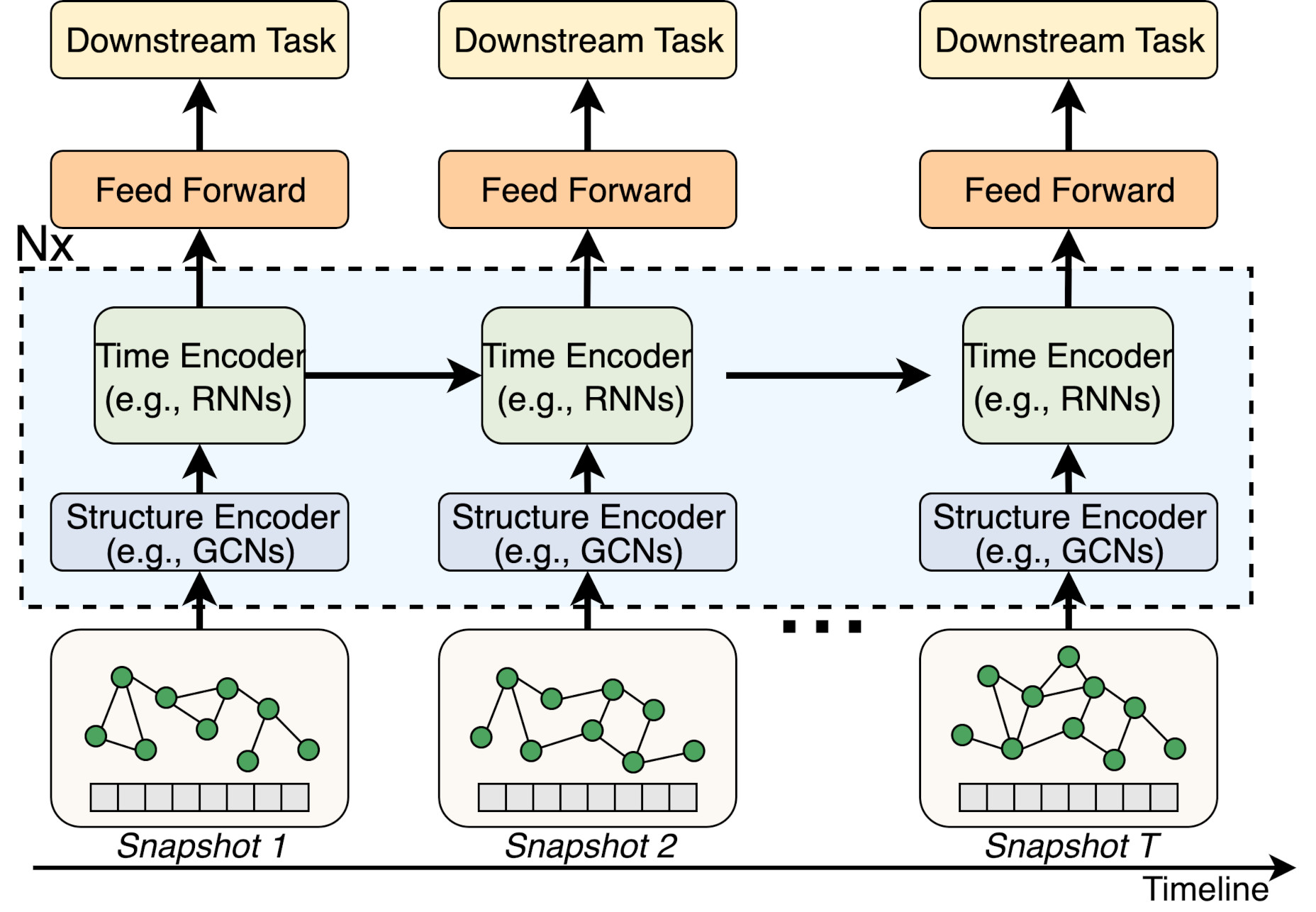}
\caption{\label{fig_dgnn}Dynamic graph neural network.}
\end{center}
\end{figure}

Graph Neural Network (GNN) has achieved great success in learning graph data in many fields, i.e., drug discovery \cite{wishart2018drugbank}, recommendation systems \cite{ ying2018graph}, and social networks \cite{wu2020graph}. Existing GNN can handle only static graphs, where vertices and edges, as well as associated features, have no change across time. However, many practical applications generate dynamic graphs whose vertices and edges change over time. Typical examples include traffic graphs that describe real-time traffic flows of roads \cite{zhang2018gaan,zhao2019t,guo2021learning}, and social networks where edges representing friend connections could be created or removed as the change of social relationship \cite{ma2020streaming, zhang2022improving}.
The strong demand of processing dynamic graphs motivates the design of the Dynamic Graph Neural Network (DGNN). As shown in Figure \ref{fig_dgnn}, a dynamic graph is divided into a number of snapshots, each of which represents the graph at a specific time. These snapshots are fed to structure encoders (e.g., GCN), respectively, followed by time encoders (e.g., RNN) to exploit temporal relationship across snapshots. By stacking multiple layers of structure encoders and time encoders, DGNN has a strong capability of capturing spatio-temporal features of dynamic graphs. Based on this basic model, various DGNN variants, EvolveGCN \cite{pareja2020evolvegcn} and DySAT \cite{sankar2020dysat}, have been proposed recently and dynamic graph learning has become a booming research area.

\begin{figure*}[t] 
\begin{center}
\includegraphics[width=0.95\textwidth]{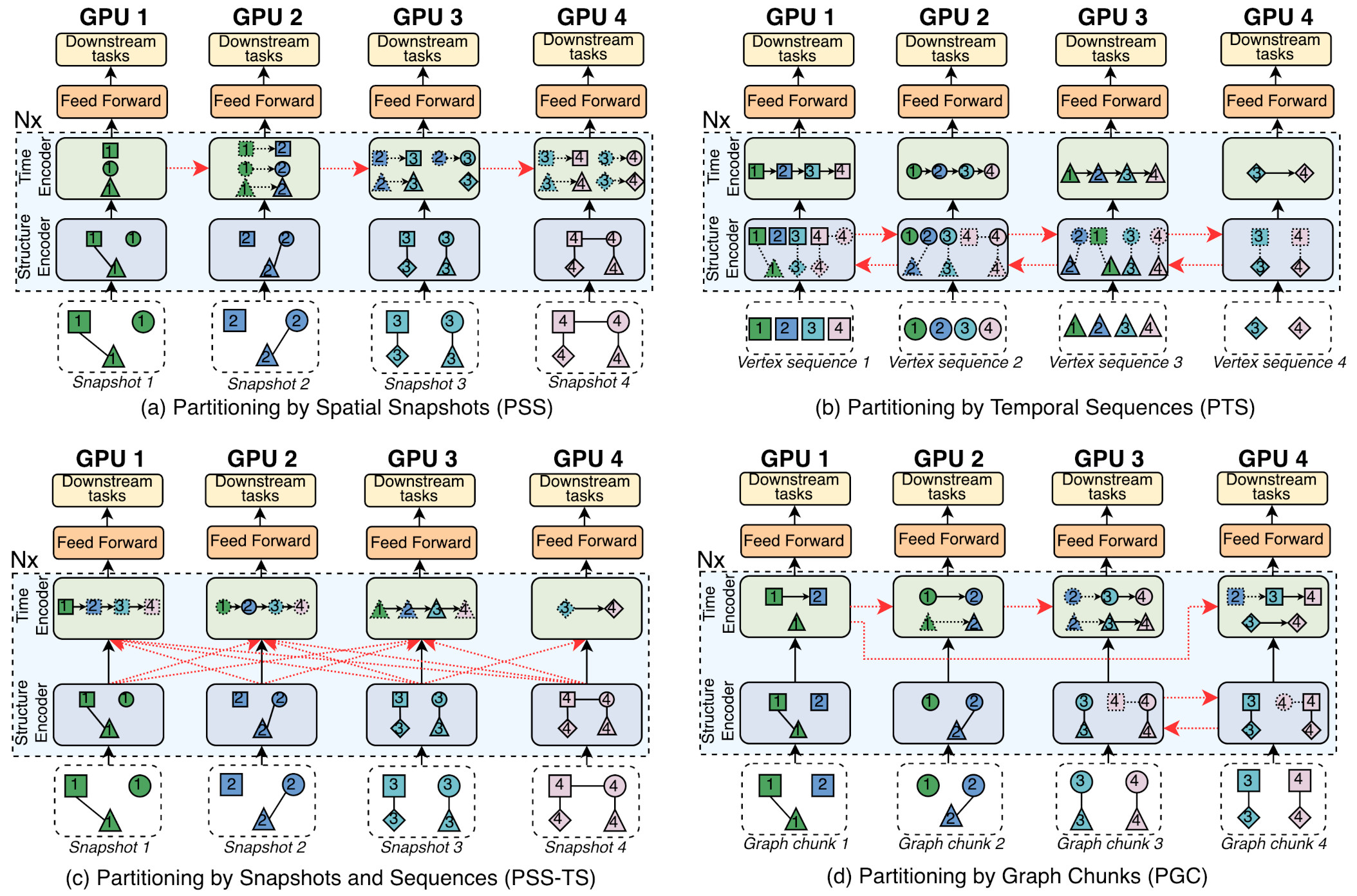}
\caption{\label{partition_motivation}Different dynamic graph partitioning methods for distributed training. Different node shapes (such as rectangles, circles, triangles, and rhombuses) represent different vertices, while colors and numbers signify vertices belonging to different snapshots. Red dotted arrows indicate communication between GPUs. Within each GPU, vertices aggregated from other GPUs are represented by dotted lines. 
}

\end{center}
\end{figure*}

Despite the great research enthusiasm for DGNN model design, its system-level support has been seldom studied. Since dynamic graphs could be very large, DGNN training usually runs on distributed systems consisting of multiple GPUs or other accelerators. To build an efficient distributed DGNN system, one of the most crucial challenges is how to partition the dynamic graph among multiple GPUs to minimize cross-GPU traffic, which has been recognized as the main system bottleneck by existing work \cite{guan2022dynagraph,chakaravarthy2021efficient}. 
A straightforward idea is to partition the dynamic graph into snapshots and assign them to GPUs, as shown in Figure \ref{partition_motivation}(a). This method is referred to as partitioning by spatial snapshots (PSS). 
However, PSS would incur high communication costs when handling dynamic graphs with long temporal information since time encoders need to share temporal embeddings across GPUs. 

To eliminate temporal embedding transmissions, the method of partitioning by temporal sequences (PTS) has been proposed \cite{guan2022dynagraph}. As shown in Figure \ref{partition_motivation}(b), PTS divides dynamic graphs into temporal sequences. Each sequence contains the same vertex's embeddings of different time, and it is the basic assignment unit to GPUs. PTS hides all communication of temporal embedding sharing within each GPU, but pays the cost of aggregating spatial embeddings across GPUs. Recently, Chakaravarthy et al. \cite{chakaravarthy2021efficient} have proposed a joint partitioning method to take the benefits of both PSS and PTS methods. As illustrated in Figure \ref{partition_motivation}(c), it applies PSS to assign snapshots to GPUs and runs structure encoders. Then, generated embeddings are shuffled by PTS, so that the ones of the same temporal sequence are gathered into the same GPU. This method is referred to as PSS-TS. Although both spatial and temporal communication is avoided, the embedding shuffling process incurs additional communication cost.

We have conducted a quantitative study (in \S\ref{motivation}) on the aforementioned methods by comparing their performance on various datasets. Our results indicate that these methods demonstrate different performance on these datasets, and there is no single method that always outperforms the others. We find that main reason of this inconsistency is an implicit assumption of these methods that dynamic graphs are uniform in both spatial and temporal structures. However, many graphs are not uniformly structured, with some snapshots being very dense while others are sparse.
Additionally, temporal sequences could have different lengths, with some vertices existing for a long time and being associated with long temporal sequences while others have short sequences. Our experiments in \S\ref{motivation} have also revealed that such spatio-temporal non-uniformity is prevalent in popular datasets. This important observation motivates us to re-examine the graph partitioning problem for DGNN training, and to design a new dynamic graph partitioning method aware of such spatio-temporal non-uniformity, so that it can always outperform existing ones on a variety of datasets. 

As a positive response to this challenging problem, we design DGC, a distributed system for efficient DGNN training, implementing a new method called partitioning by graph chunk (PGC). Different from existing works that treat snapshots or temporal sequences as basic partitioning units, we propose to partition dynamic graphs into chunks that are essentially sub-graphs across the spatial and temporal boundaries. As shown in Figure \ref{partition_motivation}(d), each graph chunk may contain vertices and edges belonging to different snapshots and temporal sequences. We design a graph chunk generation algorithm based on the graph coarsening technique with a full consideration of spatio-temporal non-uniformity, so that each graph chunk has modest training workload and few edge connections to other chunks. By a simple heuristic to assign these chunks to GPUs, DGC can achieve better workload balance and reduced communication cost, to significantly improve DGNN training efficiency. 

In addition, we propose two techniques to optimize the run-time of DGC by exploiting unique characteristics of graph chunks for further performance improvement. The first one is called chunk fusion. The graph chunks assigned to a GPU need to first go through the structure encoder. A default scheme is to load and train these chunks one by one, which would be inefficient because of redundant data loading and low GPU utilization. To address this issue, we propose to fuse these chunks into larger ones before loading, while considering the GPU memory constraint. Furthermore, the temporal sequences sent to the time encoder could have different lengths. In order to pack them for GPU processing, we need to align these sequences by padding a large number of zeros, which could waste GPU memory. Thus, we propose to fuse these sequences by concatenating short ones to reduce padded zeros. However, an intuitive sequence concatenation scheme would generate incorrect outputs of time encoders, and thus impose negative influence on training accuracy. We design a masking scheme for the time encoder, so that it can generate correct embeddings while padded zeros can be reduced.

Second, we propose adaptive stale embedding aggregation to further reduce communication cost among GPUs. This is motivated by the observation that vertices may generate similar embeddings in different training epochs (\S\ref{stale_agg_sec}). DGC allows GPUs to reuse stale embeddings from previous epochs if they are sufficiently similar, to reduce data traffic between GPUs. However, using stale embeddings could slow down the training convergence or even decrease the final training accuracy. It is quite challenging to estimate embedding similarity and decide when they can be reused, to balance communication cost and training convergence. We propose an adaptive stale aggregation scheme, which decides whether stale embeddings could be used according to the current training loss.





We deploy DGC on an 8-GPU testbed and conduct experiments using four different dynamic graphs and three representative DGNN models (including T-GCN \cite{zhao2019t}, DySAT \cite{sankar2020dysat}, and MPNN-LSTM \cite{malik2021dynamic}). The experimental results show that DGC achieves a 1.25$\times$ - 7.52$\times$ speedup over the state-of-the-art. We also conduct ablation experiments to study the benefits of our proposed run-time optimization techniques.


The rest of this paper is organized as follows. We present the preliminary and the motivation in \S\ref{sec_pre_mot}. A system overview is in \S\ref{system_overview}. We present the method of partitioning by graph chunk in \S\ref{pgc_sec}, followed by the run-time optimizations in \S\ref{run_time_opt}. \S\ref{implementation} discusses the implementation and \S\ref{evaluation} presents our experimental results. Related work is in \S\ref{related_work}. \S\ref{conclusion} finally concludes this paper.

\section{Preliminaries and Motivations} \label{sec_pre_mot}

\subsection{Preliminaries} \label{preliminaries}



\textbf{Dynamic Graphs. }A dynamic graph can be represented by $\mathcal{G}=\{G_{1}, G_{2},..., G_{T}\}$, where $G_{t}=(V_{t}, E_{t})$ is a snapshot at timestep $t$. $V_{t}$ and $E_{t}$ represent the vertex and edge sets of snapshot $G_{t}$, respectively. Each vertex $v_{i,t}$ in $V_{t}$ is associated with a feature vector $x_{i,t}$. In addition, a vertex $v_{i,t}$ has spatial and temporal neighbors. (1) \textit{Spatial neighbors}, denoted as $\mathcal{NS}(i,t)$, are the vertices that are directly connected to $v_{i,t}$ through an edge in the same snapshot $G_{t}$. They represent the immediate connections or relationships among the vertices in a specific timestep. (2) \textit{Temporal neighbors}, denoted as $\mathcal{NT}(i,t)$, are the vertices corresponding to the same entity as $v_{i,t}$ but in different snapshots. They represent the changes or evolution of vertex features across different timesteps.

\noindent\textbf{Dynamic Graph Neural Networks}. A dynamic graph neural network (DGNN) is composed of multiple blocks, where each block consists of a \textit{structure encoder} and a \textit{time encoder}, as illustrated in Figure \ref{fig_dgnn}. The structure encoder extracts hidden information for each vertex by aggregating information from its structural neighbors. Meanwhile, the time encoder accumulates information for each vertex from its temporal neighbors. Note that different DGNN models have different implementation of structure and time encoders. For example, T-GCN \cite{zhao2019t} uses three 2-layer GCN \cite{kipf2016semi} as the structure encoder, and a 1-layer GRU \cite{cho2014learning} model as the time encoder. DySAT \cite{sankar2020dysat} incorporates a 1-layer graph attention network (GAT) \cite{velivckovic2017graph} and a 1-layer scaled dot-product attention model \cite{vaswani2017attention} within each of its DGNN blocks.


\noindent\textbf{Distributed DGNN training}.
Distributed DGNN training across multiple GPUs is a promising approach for handling large dynamic graphs. However, the challenge lies in determining how to partition the dynamic graph. The partitioning algorithm should minimize cross-GPU communication by reducing data dependency breakdown while maintaining workload balance.

Graph partitioning has been extensively studied in distributed GNN training for static graphs \cite{ma2019neugraph, jia2020improving, cai2021dgcl, md2021distgnn, gandhi2021p3, wang2021gnnadvisor}. However, these methods designed for unraveling spatial dependency cannot be applied to dynamic graph partitioning with complex temporal dependency, which motivates several recent works about dynamic graph partitioning. These existing works can be classified into three categories. (1) Partitioning by Spatial Snapshots (PSS): it treats a snapshot as the partition unit and always keeps spatial dependencies within the same GPU. (2) Partitioning by Temporal Sequences (PTS): A temporal sequence records the states of the same vertex in different time. This approach eliminates communication overhead for vertices when aggregating their temporal neighborhoods. However, high communication overhead may arise when vertices aggregate their spatial neighborhoods, as spatial dependencies are broken down across GPUs. (3) Partitioning by Snapshots and Sequences (PSS-TS): a joint method adopts PSS for the structure encoder while transitioning to PTS for the time encoder. This approach avoids communication overhead when aggregating both spatial and temporal neighborhoods. However, it involves additional shuffling cost to re-assign vertices across GPUs.

\begin{table}[]
\begin{center}
\begin{tabular}{l|c c c c}%
\hline
\textbf{Attributes} & Amazon & Epinion & Movie & Stack\\
\hline
\textbf{\# of snapshots} & 121 & 500 & 289 & 93\\
\textbf{Total \# of vertices} & 103M & 72M & 43M & 83M\\
\textbf{Total \# of edges} & 5.7M & 13M & 27M & 47M\\
\hline
\end{tabular}
\caption{\label{Dataset} Dynamic Graph Datasets.}
\end{center}
\end{table}

\begin{figure}[t]
\subfigure[\label{str_cdf}CDF of number of spatial neighbors.]{
\centering
\includegraphics[width=0.22\textwidth]{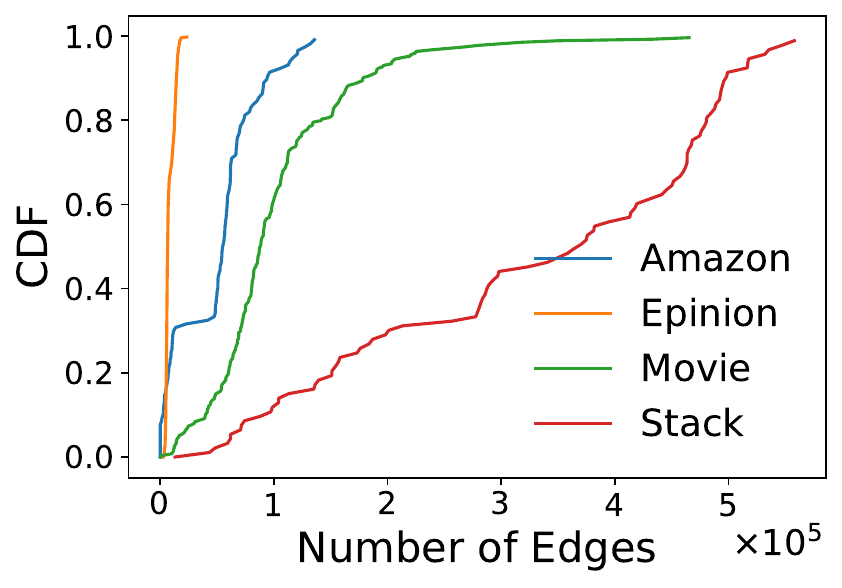}}
\subfigure[\label{tem_cdf}CDF of vertex sequence lengths.]{
\centering
\includegraphics[width=0.22\textwidth]{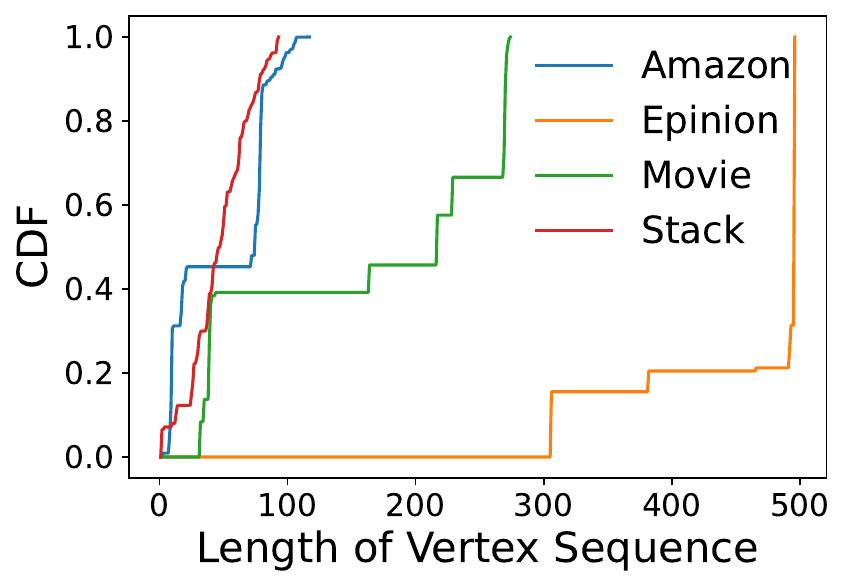}}
\centering
\caption{Spatio-temporal non-uniformity of different datasets.}
\label{Stack_cdf}
\end{figure}

\subsection{Motivation} \label{motivation}

\subsubsection{Spatio-temporal non-uniformity of dynamic graphs} 
We demonstrate the spatio-temporal non-uniformity of dynamic graphs using four datasets, whose details are shown in Table \ref{Dataset}. In Figure \ref{str_cdf}, we plot the cumulative distribution function (CDF) curves of the number of edges within snapshots. We observe that some snapshots have very few edges, while others could be very dense, indicating diverse spatial features among snapshots. The CDF about lengths of vertex sequences is shown in Figure \ref{tem_cdf}. Some vertices exist for a long time and thus have long temporal sequences, while others are short. 



\subsubsection{Performance of dynamic graph partitioning methods on different datasets}\label{motivation_1}

\begin{figure}[t]
\subfigure[\label{time_break_down}Average epoch time.]{
\centering
\includegraphics[width=0.45\textwidth]{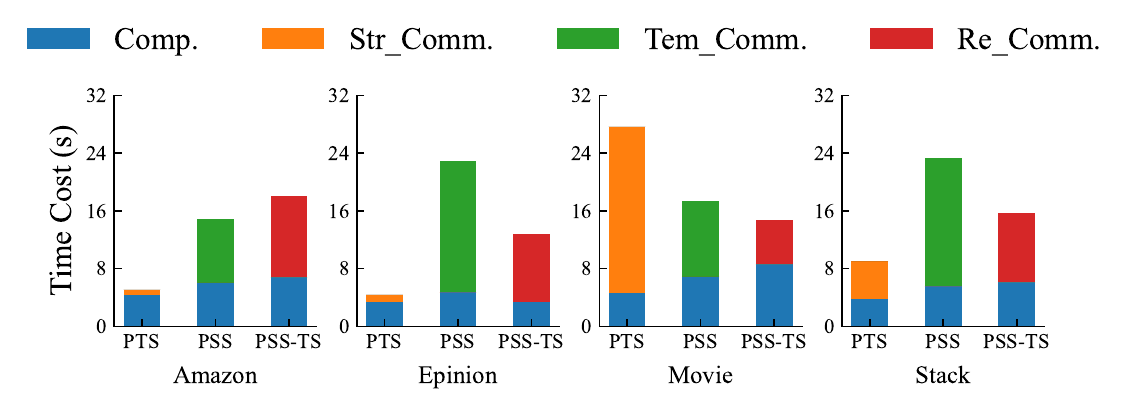}
}
\quad
\subfigure[\label{workload_div}GPU workload divergence.]{
\centering
\includegraphics[width=0.45\textwidth]{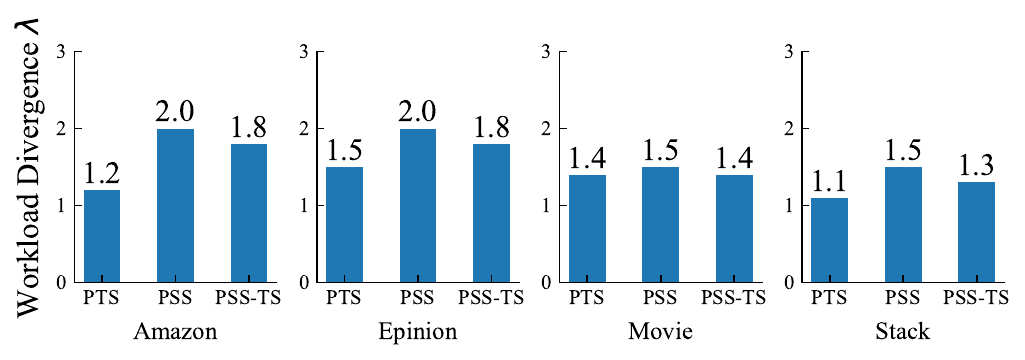}}
\caption{\label{analysis}Performance of dynamic graph partitioning methods on different datasets.}
\centering
\end{figure}

We use the four datasets in Table \ref{Dataset} to train DySAT \cite{sankar2020dysat} models on 4 NVIDIA V100 GPUs. 
The average epoch time of PSS, PTS, and PSS-TS is shown in Figure \ref{time_break_down}.
Although all methods are performing the same training task using the same dataset and model, they are different in graph partitioning and thus assign different workloads to GPUs.
We can see that the PTS has the shortest epoch time on Amazon, Epinion, and Stack datasets, but longer than PSS and PSS-TS methods on the Movie dataset. The breakdown of computation time and communication time is also shown in this figure. For the Epinion dataset, all methods have similar computation time, but PSS has much longer communication time, because more nodes are involved in temporal computation and their embeddings are shared across GPUs. However, since the Movive dataset has dense spatial structures, PTS breaks this structure and incurs higher communication overhead. Although PSS-TS avoids communication overhead within both spatial and temporal computations, it incurs significant overhead because of embedding re-assignment, especially when the number of vertices is large (e.g., Amazon and Stack datasets).
The fact in Figure \ref{time_break_down} demonstrates that different datasets show distinct spatio-temporal features and neither method can always win over all datasets.

We also conduct experiments to study GPU load balance of existing methods. For the PSS method, we follow the strategies proposed in \cite{guan2022dynagraph, chakaravarthy2021efficient} to assign the same number of snapshots to each GPU. Similarly, we let each GPU get the same number of sequences in the PTS method. We define a metric $\lambda=\frac{T_{max}}{T_{min}}$ to evaluate the level of GPU load balance, where $T_{max}$ and $T_{min}$ is the maximum and minimum epoch time, respectively, among GPUs. If the value of $\lambda$ is close to 1, training workloads are well balanced. Otherwise, faster GPUs need to wait for slower ones, leading to low hardware utilization. As shown in Figure \ref{workload_div}, we find that PTS method has a good load balance with $\lambda=1.1$ under the Stack dataset, but its load balance becomes worse when training other datasets. PSS and PSS-TS have bigger $\lambda$, indicating stronger imbalance of training workloads among GPUs.

\subsubsection{Performance of dynamic graph partitioning methods within a single dataset}
\begin{figure}[t]
\subfigure[\label{part_node_pdf}Number of vertices on different parts.]{
\centering
\includegraphics[width=0.22\textwidth]{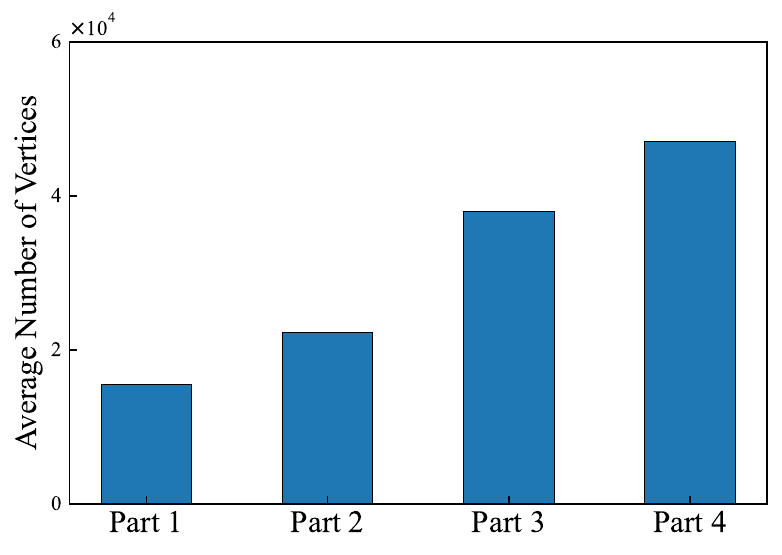}}
\subfigure[\label{part_edge_pdf}Number of edges on different parts.]{
\centering
\includegraphics[width=0.22\textwidth]{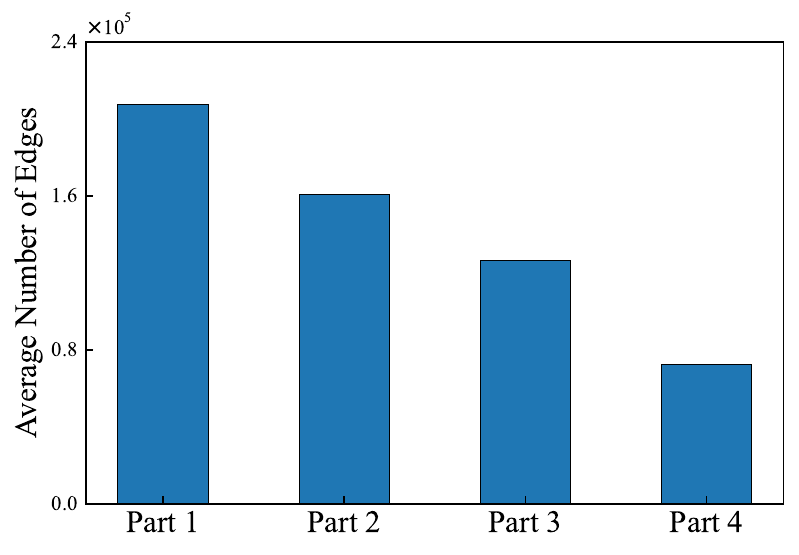}}
\quad
\subfigure[\label{Movie_part_analy}Training time breakdown on Movie.]{
\centering
\includegraphics[width=0.45\textwidth]{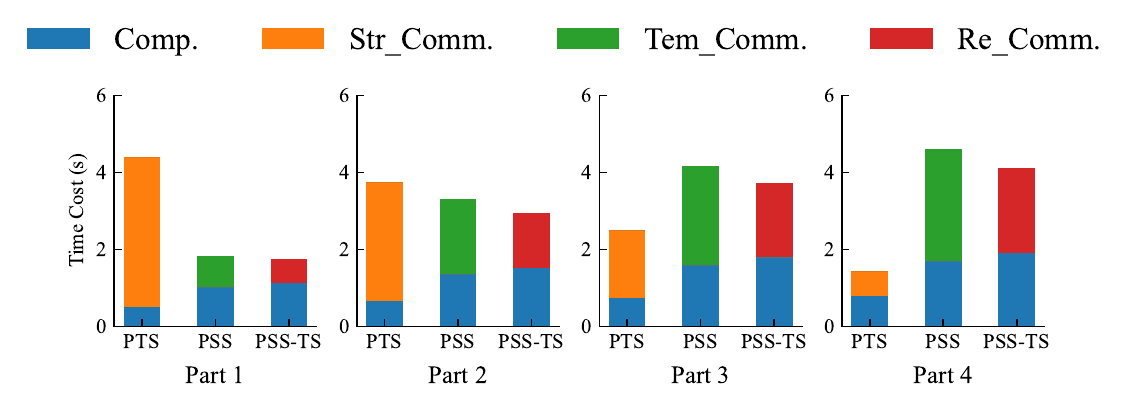}}
\centering
\caption{\label{in_graph_analysis}Performance of dynamic graph partitioning methods within a single dataset.}
\end{figure}
We then dig into the Movie dataset to study its internal spatio-temporal features. We divide the whole dataset into 4 parts by snapshots, and each part has the same number of snapshots. The average number of vertices and edges in each part are shown in Figure \ref{part_node_pdf} and Figure \ref{part_edge_pdf}. We can see that the number of vertices changes significantly across different parts. The fourth part has 3 times more vertices than the first one. Meanwhile, the number of edges also changes, but with a different pattern from vertices. For example, the first part has the most edges but it has only half of vertices of the fourth part. 

To study how internal spatio-temporal features affect graph training, we measure the epoch time of three methods on 4 sub-datasets. As shown in Figure \ref{Movie_part_analy}, these methods show distinct performance. PSS and PSS-TS outperform PTS when training the first and second sub-datasets, thanks to its much shorter communication time. However, PTS has better performance on the third and fourth sub-datasets. We also find that the communication overhead of PSS and PSS-TS grows as the increasing of number of vertices. In contrast, PTS has longer epoch time on sub-datasets with more edges. That is because PSS partitions dynamic graph data by snapshots, and denser snapshots incur more data sharing over networks. PTS conducts data partition by sequences, i.e., cutting edges within snapshots, and thus more edges would generate more traffic. 

We also conduct experiments for other 3 datasets and have similar observations. For even a single dataset, its different parts show distinct spatio-temporal features.
However, existing works are unaware of such graph internal diversity and apply a single partitioning strategy for the whole dataset.

\section{System Overview} \label{system_overview}

\begin{figure}[t] 
\begin{center}
\includegraphics[width=0.45\textwidth]{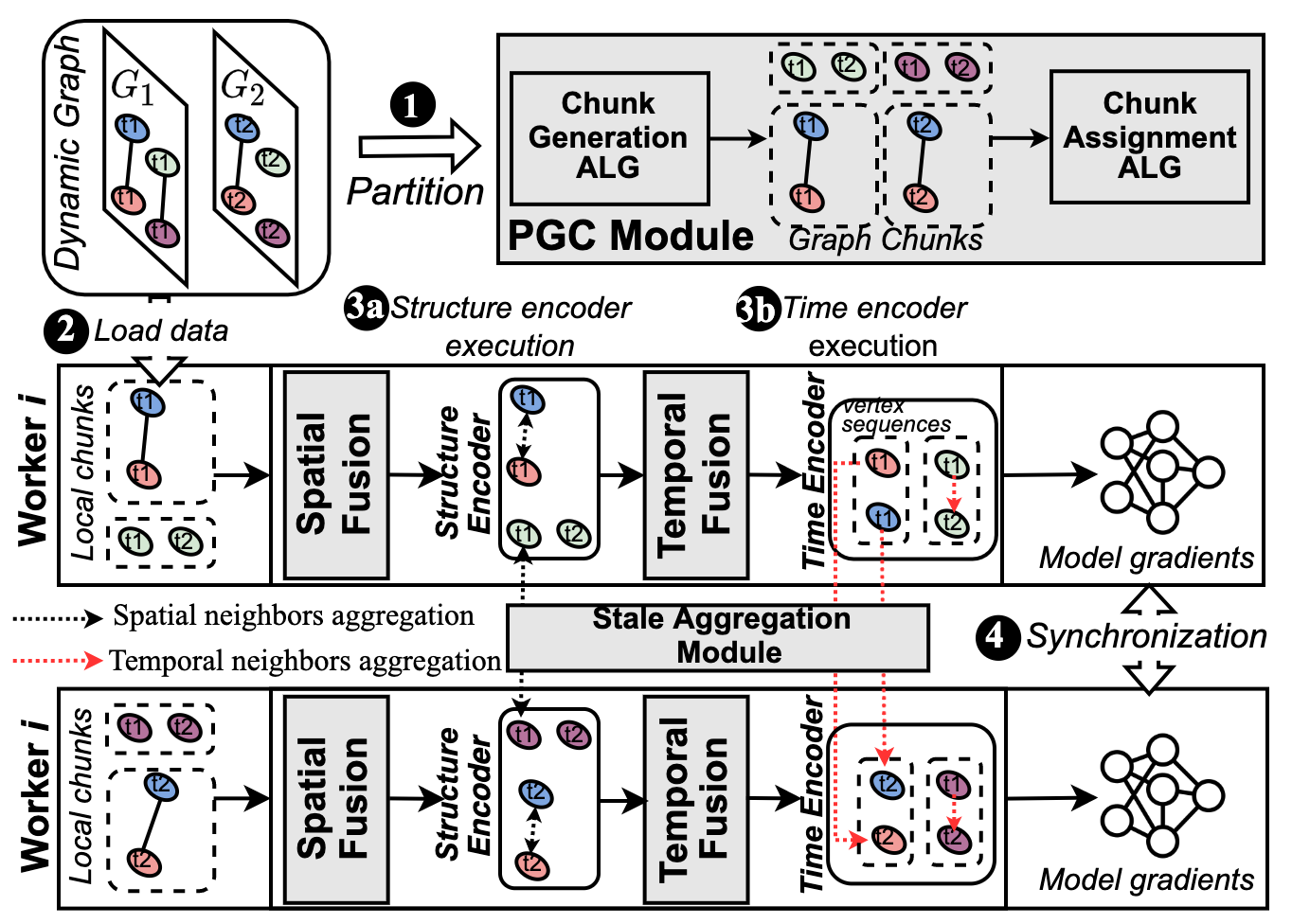}
\caption{\label{overview}System Overview. DGC contains an offline PGC module and an optimized run-time. The PGC module partitions the dynamic graph into chunks and assigns them to GPUs. The run-time contains two new modules of chunk fusion and adaptive stale embedding aggregation.}
\end{center}
\end{figure}

In this section, we present an overview of the DGC design. 
We first set our design goals as follows.

\noindent\textbf{High training efficiency}. 
Due to massive data dependencies (including spatial and temporal dependencies) among vertices in dynamic graphs, distributed DGNN training suffers from high communication cost that would be the performance bottleneck. DGC needs to reduce communication cost to accelerate training process.

\noindent\textbf{High GPU utilization}. Multiple GPUs are used to train DGNNs for handling large dynamic graphs. GPU utilization is a crucial metric for efficient resource management. DGC should ensure high GPU utilization during DGNN training.

\noindent\textbf{Consistent training convergence}. While introducing various optimizations to accelerate DGNN training, DGC needs to ensure that these designs do not compromise training convergence, preserving the quality of the final model.

Figure \ref{overview} illustrates an overview of DGC's design. In order to handle large dynamic graphs, DGC uses multiple GPUs that collaboratively train the DGNN model. Specifically, DGC maintains some training workers, and each worker is bonded to a GPU. The system workflow is as follows. 
\ding{182} First, a PGC (Partitioning by Graph Chunk) module partitions the dynamic graph into multiple graph chunks and assigns them to workers. \ding{183} Based on the partitioning and assignment results, each GPU worker loads their assigned graph chunks, and then trains the corresponding DGNN model for multiple epochs. As mentioned in \S\ref{preliminaries}, a DGNN model could include multiple blocks, and each block consists of a structure encoder and a time encoder. In Figure \ref{overview}, we show a DGNN model with a single block for simplicity. 
\circled{3a} The structure encoder computes spatial embeddings of local vertices. Note that spatial neighbors may be located on other workers, their embeddings need to be transmitted over the network. \circled{3b} Similarly, time encoders of different workers need to share embeddings of temporal neighbors. Since these embeddings could be very large, frequent embedding sharing would incur high communication costs.
\ding{185} Finally, each GPU worker calculates gradients based on a loss function and synchronizes them with other workers, so that they can update DGNN model weights and proceed to the next training epoch.

We can see that DGC is different from traditional data parallelism and model parallelism, which are popular distributed training approaches adopted by CNN or transformer models, because of the complicated spatio-temporal dependency. The whole system performance is mainly affected by communication costs among GPUs and their workload balance. In order to achieve our design goals, we design the following three key modules.

\noindent
\textbf{PGC:} The PGC module partitions the dynamic graph, by introducing the concept of graph chunks to minimize cross-GPU communication cost while maintaining workload balance (\S\ref{pgc_sec}).

\noindent
\textbf{Chunk fusion:} This module fuses multiple chunks assigned to each GPU to reduce data loading cost, so that the GPU utilization can be significantly improved (\S\ref{chunk_fusion_sec}).

\noindent
\textbf{Adaptive stale embedding aggregation:} We observe that some vertex embeddings have no big changes in different training epochs. Thus, we are motivated to propose a stale aggregation module that enables some GPUs to reuse some previously received embeddings if there is only trivial difference. Many embedding transmissions can be avoided to reduce communication costs (\S\ref{stale_agg_sec}).

\section{Partitioning by Graph Chunks} \label{pgc_sec}

The issues of PTS and PSS stem from their high-level semantic graph partitioning, i.e., in units of snapshots or sequences, without considering the potential influence to running efficiency of the distributed system. Since snapshots and temporal sequences could be very large, it leaves little optimization space for the following workload assignment algorithm. No matter how sophisticated assignment algorithms are designed, it is still difficult to achieve good workload balance among GPU while minimizing cross-GPU traffic.

To fundamentally solve these issues, we propose the method of partitioning by graph chunk (PGC), by jointly considering graph features and hardware resources. PGC partitions graphs into chunks that are sub-graphs across spatial and temporal boundaries of original dynamic graphs. Each graph chunk has modest training workload and few edge connections to other chunks, so that even a simple chunk assignment algorithm can achieve significant efficiency improvement.

However, designing an efficient PGC is challenging. Since the weaknesses of PTS and PSS are mainly because of their coarse-grained partition at the snapshot or sequence level, a straightforward improvement is to treat dynamic graphs as a super graph by linking vertices with temporal relationship. We can then partition this super graph into several parts, each of which is assigned to a GPU for training. Such a kind of graph partitioning has been widely studied by existing works \cite{stanton2012streaming, tsourakakis2014fennel, zhang2017graph}. Even though it is an NP-hard problem, there exist methods with good theoretical and empirical results. However, these methods have high computational overhead, which can be hardly applied for dynamic graphs with millions or even billions of vertices and edges. We address this overhead challenge by borrowing the idea of graph coarsening \cite{purohit2014fast, lasalle2015improving, bravo2019unifying, huang2021scaling}, and customizing it for dynamic graph partitioning. 
In addition, we design a fast algorithm to assign chunks to GPUs. 

\subsection{Chunk Generation} \label{chunk_gen_sec}
The chunk generation algorithm is based on label propagation and its basic idea is to assign a unique label to each vertex, which is then propagated along graph edges and be updated iteratively according to a label updating policy. Finally, vertices with the same label can be grouped together to form a chunk.

Two key challenges must be addressed to make this algorithm work efficiently for dynamic graphs. First, traditional label propagation is constrained within snapshots (because there is no edge between snapshots) and temporal features cannot be fully exploited. Therefore, we add virtual temporal edges between temporal vertices so that labels can be propagated across snapshots.

Second, even with these virtual edges, the label propagation algorithm could be difficult to generate chunks with minimum inter connections as we desire, because the algorithm lets labels have the same opportunity to travel along all edges. However, spatial edges and temporal edges have different communication cost. For example, T-GCN involves two GCN layers and one GRU layer for each DGNN block, which means that vertices aggregate their spatial neighborhoods twice, while only aggregating temporal neighborhoods once. To reflect this unique characteristic, we propose to customize edge weights during label propagation according to their communication cost. Specifically, we initialize the label of each vertex $v_{i,t}$ as follows:
\begin{align}
c(v_{i,t}) = \sum_{\tau=1}^{t-1}|V_{\tau}| + i,
\end{align}
where $|V_{\tau}|$ is the number of vertices in snapshot $G_{\tau}$ and $\tau\in [1, t-1]$, so that each vertex can get a unique label. After initialization, the algorithm runs several iterations of label propagation. In each iteration, vertices propagate labels to both spatial and temporal neighbors and update their labels in a \textit{weighted} manner. Specifically, each vertex $v_{i,t}$ receives multiple labels from its neighbors, and these labels are maintained in a set $\mathcal{L}(v_{i,t})$. The set of vertices sending the same label $c$ is denoted by $S(c)$. Each label $c$ is associated with a weight $weight(c)$ that is the total amount of traffic for embedding sharing from vertices in $S(c)$ to $v$.
Note that $weight(c)$ may vary depending on the specific DGNN model and can be easily obtained through profiling. Vertex $v_{i,t}$ updates its label by:
\begin{align}
    c(v_{i,t}) = \argmax_{c\in \mathcal{L}(v_{i,t})} weight(c),
\end{align}
which chooses the label with the maximum weight. The rationale is as follows. Recall that our final goal is to create graph chunks with minimum inter connections. Since directly minimizing inter-chunk connections could be difficult, we convert the problem into an equivalent one of maximizing the communication cost within chunks. The equivalence can be proved by formulating both problems and showing that the sum of their objective functions is a constant, i.e., the total cost of all edges. Therefore, we choose a neighboring label with the most weight, so that they can be grouped together as a chunk. The above process is repeated until convergence, i.e., no labels can be changed. Note that we control the maximum size of chunks by constraining the propagation of some labels if they are attached to too many vertices.

\noindent\textbf{Discussion}. To maximize GPU utilization, an alternative method is to let graph chunks expand until they reach GPU memory capacity during chunk generation. 
However, due to the convergence of label propagation, this method cannot guarantee that each generated chunk perfectly saturates GPU memory. Imposing chunks to expand to GPU memory would falsely group vertices, leading to high cross-GPU traffic. Our design respects the convergence of label propagation and uses a fast algorithm to fuse chunks (in \S\ref{sec_sf}) for high utilization of GPU memory.

\begin{figure}[t] 
\begin{center}
\includegraphics[width=0.48\textwidth]{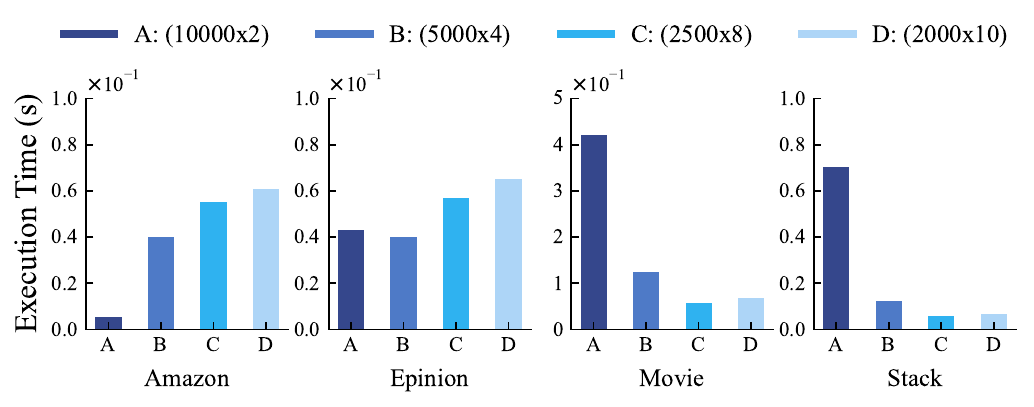}
\caption{\label{chunk_workload_eva}The execution time of different chunks with the same number of vertices.}
\end{center}
\end{figure}

\subsection{Chunk Assignment} \label{chunk_assignment_sec}
After chunk generation, we need to assign generated chunks to GPUs, by considering cross-GPU communication cost and workload balance among GPUs. An important step in chunk assignment algorithm design is to evaluate chunk workload. A simple and straightforward method of evaluating chunk workload is to count the numbers of vertices and edges \cite{lin2020pagraph}. However, this method cannot provide sufficient estimation accuracy because the execution time of a chunk on a GPU is determined by many factors, such as the number of vertices, number of edges, sequence lengths, feature dimension, and others. Our experiments have also confirmed this point. As shown in Figure \ref{chunk_workload_eva}, we generate four chunks with the same number of vertices, but their execution time is different and the maximum gap could be 8 times. 

To address this challenge, we propose a learning-based method to accurately evaluate the workload of each chunk. Specifically, we use multi-layer perceptions (MLPs) to predict the execution time of training operations associated with a chunk as its workload. Furthermore, we find that there are two kinds of operations, e.g., spatial ones and temporal ones, which consume different time. Therefore, we train two separate MLPs to predict the execution time of structure and time encoders, respectively. More details about the implementation details of prediction MLPs are given in \S\ref{implementation}. 

\begin{algorithm}[t]
\caption{Chunk Assignment Algorithm}
\label{chunk_assign_alg}
\begin{algorithmic}[1]
\Require A set of graph chunks $A$, and a set of GPUs $M$
\Ensure Assignment decisions: $x_{a}$, $\forall a \in A$, denoting the GPU of chunk $a$;
\State A set of chunks already assigned to GPU $m$: $Q_{m} \leftarrow \varnothing$, $\forall m\in M$;
\State Profile workloads of chunks with MLPs, denoted by $g_{a}$; \label{profile}
\State Sort chunks in decreasing order of $g_{a}$, as $\tilde{A}$; \label{sort}
\For{$a \in \tilde{A}$}
    \For{$m\in M$}
        \State 
        \begin{equation}
            s_{m} = (\overline{g} - \sum_{a'\in Q_{m}}g_{a'})\cdot \sum_{a'\in Q_{m}}h(a, a');\label{score_m}
        \end{equation}
    \EndFor
    \State $m^{*} = \argmax_{m}s_{m}$;
    \State $x_{a} = m^{*}$, $Q_{m^{*}}.append(a)$;
\EndFor
\end{algorithmic}
\end{algorithm}

We design a heuristic algorithm to assign generated chunks to GPUs, whose pseudocodes are shown in Algorithm \ref{chunk_assign_alg}. We first predict the workload of each chunk $a$ using the proposed MLPs. Chunks are sorted in decreasing order of their predicted workloads. Then, for each chunk $a$, we compute assignment scores, which are defined in Eq. (\ref{score_m}), for all available GPUs. The score consists of two parts. The first part $(\overline{g} - \sum_{a'\in Q_{m}}g_{a'})$ indicates the workload balance among GPUs, where $\overline{g}$ is the average workload and $Q_{m}$ is the set of chunks assigned to GPU $m$. The second part $\sum_{a'\in Q_{m}}h(a, a')$ is the communication cost between the chunk $a$ and the ones already assigned to GPU $m$.


\section{Run-time Optimization} \label{run_time_opt}

After chunk assignment, the whole training system is ready for running. In this section, we propose two techniques to optimize DGC run-time to accelerate the training process. 

\subsection{Chunk Fusion} \label{chunk_fusion_sec}
Each GPU is assigned by a number of graph chunks. A default approach is to load and process these chunks one by one, which would lead to substantial redundant data loading and low GPU utilization. 

\subsubsection{\label{sec_sf}Spatial Fusion} 

\begin{figure}
    \centering
    \includegraphics[width=0.45\textwidth]{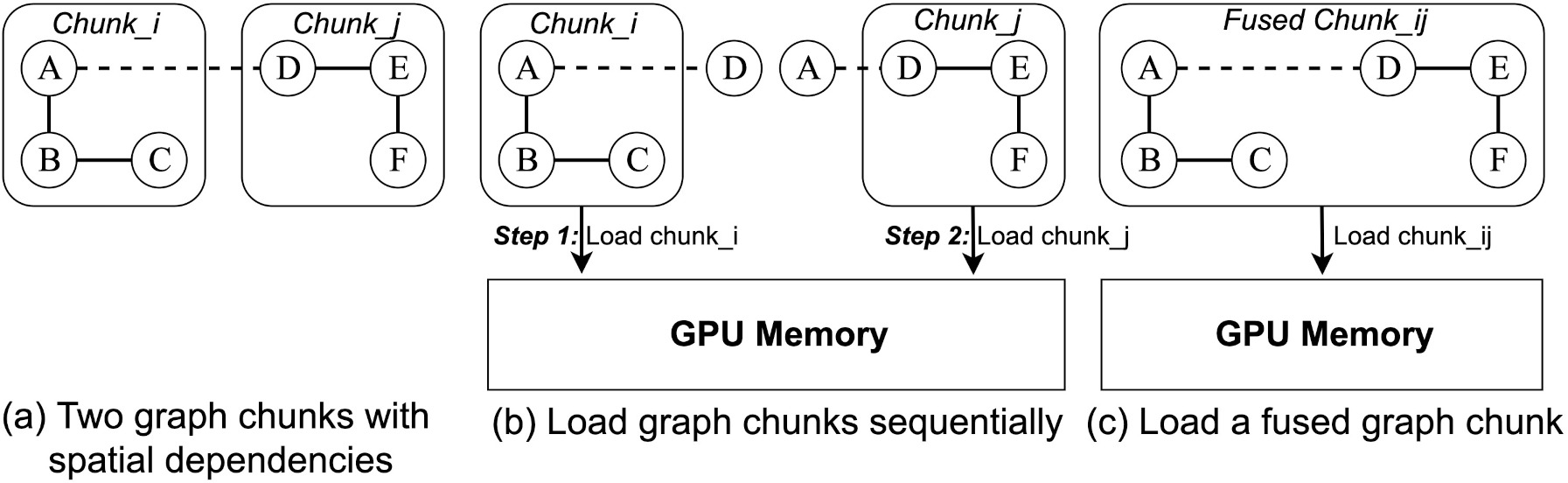}
    \caption{An illustration of spatial fusion.}
    \label{chunk_fusion_motivation}
\end{figure}

We use an example in Figure \ref{chunk_fusion_motivation} to show the motivation of spatial fusion. Two graph chunks with cross-chunk spatial dependency (e.g., the edge between $A$ and $D$) need to be loaded into the GPU for training. In a default scheme, when loading each chunk, we need to load not only the vertices within this chunk but also the ones of other chunks. For example, in Figure \ref{chunk_fusion_motivation}(b), when processing chunk\_i, we must load vertex $D$ from chunk\_j because vertex $A$ requires information from $D$. Similarly, when processing chunk\_j, we also need to load vertex $A$. Consequently, vertices $A$ and $D$ are loaded twice. 

DGC introduces spatial fusion to reduce data loading cost and improve GPU utilization. By fusing multiple chunks together, we can load them simultaneously and only load the vertices with cross-chunk dependency once, as illustrated in Figure \ref{chunk_fusion_motivation}(c). Moreover, fused chunks can be executed together to fully utilize GPU resources. 

Although spatial fusion can significantly increase GPU utilization, a GPU could be assigned with a large number of graph chunks and fusing all chunks may exceed the GPU memory limit. Therefore, we propose a simple yet effective heuristic algorithm to select a subset of graph chunks for fusion with respect to the GPU memory constraint. 
Specifically, we estimate the potential redundant data loading, in terms of the amount of data transmission, among chunks and then iteratively fuse two chunks with the maximum data transmission. When the size of any fused chunk is close to GPU memory limit, we stop to fuse them any more.  


To estimate the GPU memory consumption of a chunk, we exploit the observation that memory consumption remains relatively consistent across training epochs for each chunk. As a result, it is sufficient to determine the GPU memory consumption through a single execution. Specifically, right after completing the first training epoch, we monitor the memory usage during the training process to assess the memory consumption of each chunk. 

\begin{figure}[t]
    \centering
    \includegraphics[width=0.45\textwidth]{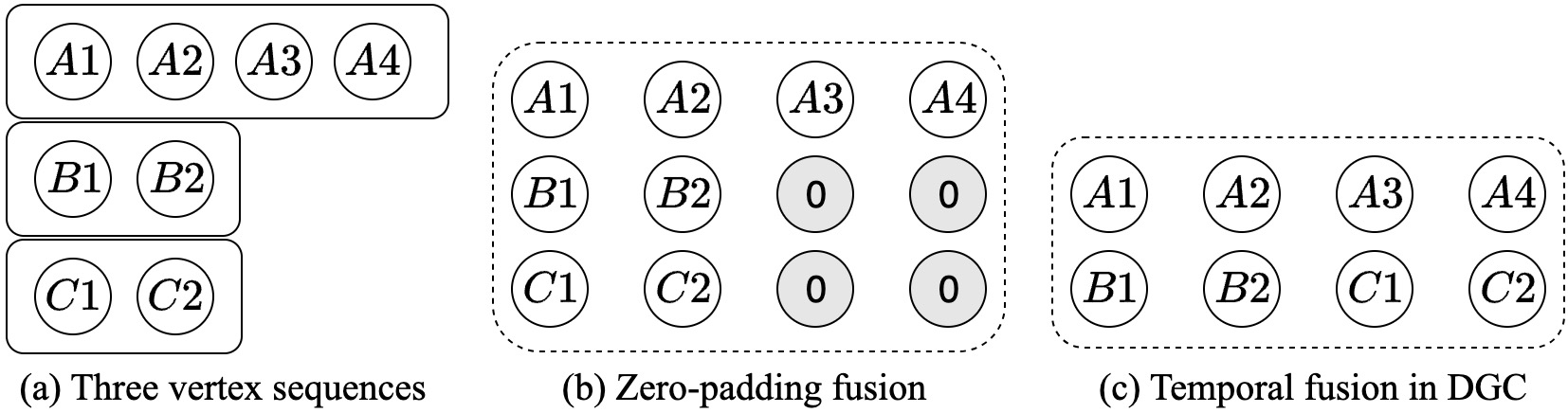}
    \caption{An illustration of temporal fusion.}
    \label{temporal_fusion}
\end{figure}

\begin{figure}[t]
    \centering
    \includegraphics[width=0.45\textwidth]{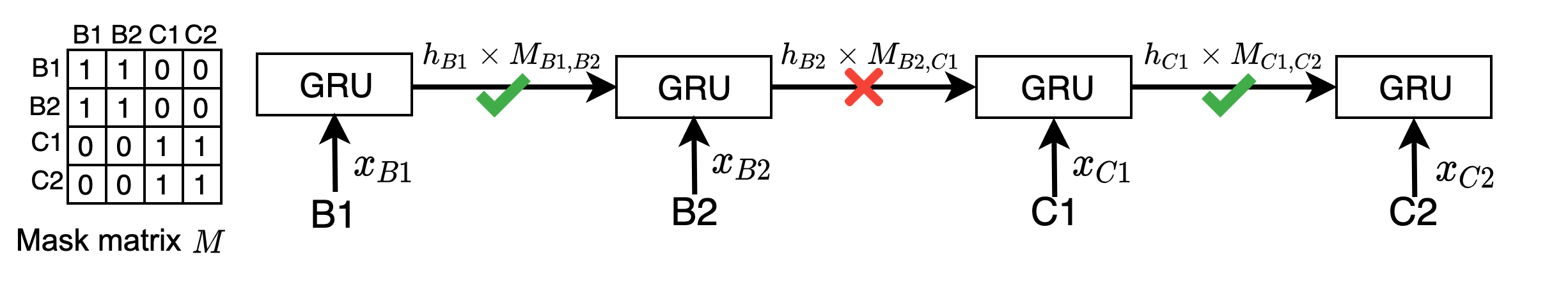}
    \caption{The mask operation guarantees the correctness of temporal fusion. We take a GRU layer as an example, which is adopted in the time encoder of T-GCN. The sequence is $(B1, B2, C1, C2)$, given in Figure \ref{temporal_fusion}(c). }
    \label{mask}
\end{figure}

\subsubsection{Temporal Fusion} 
By carefully examining the input of time encoders, we find that vertex sequences have varying lengths, as an example shown in Figure \ref{temporal_fusion}(a). A common practice is to pad zeros, so that they can be packed together and processed by GPUs, as shown in Figure \ref{temporal_fusion}(b). However, padding zeros incurs redundant computation cost and wastes GPU memory. To further increase GPU utilization, we propose to fuse embeddings by concatenating shorter sequences instead of padding zeros. As illustrated in Figure \ref{temporal_fusion}(c), we concatenate sequences $(B1,B2)$ with $(C1,C2)$ to form a new sequence of length 4, enabling simultaneous processing with another sequence $(A1,A2,A3,A4)$.

Although this method can improve GPU utilization by avoiding zero padding, the time encoder may generate incorrect output due to unnecessary message passing between vertices belonging to different sequences. For instance, if we concatenate $(B1, B2, C1, C2)$ as one sequence, vertex $C1$ receives an unwarranted hidden state from $B2$, resulting in wrong outputs. To address this issue, DGC uses a mask to ensure correct output, as shown in Figure \ref{mask}. 
Taking the sequence $(B1, B2, C1, C2)$ as an example, we calculate the update of $C1$ in time encoder of T-GCN as follows:
\begin{align}
    &u_{C1} = \sigma(W_{u}h_{B2}M_{B2,C1} + W_{u}x_{C1}+b_{u}),\\
    &M_{B2,C1}=
    \left\{\begin{matrix}
    1, \text{if }B2, C1\text{ belong to a sequence},\\
    0, \text{otherwise},
    \end{matrix}\right.
\end{align}
where $W_{u}$ and $b_{u}$ are learnable weights and bias in the update gate, respectively. The term $h_{B2}$ represents the hidden state of $B2$, and $\sigma$ is the activation function. We use the mask $M_{B2,C1}$ to prevent the hidden state of $B2$ from being added to the update gate output of $C1$.

\subsection{Adaptive Stale Embedding Aggregation} \label{stale_agg_sec}

		

\begin{figure}[t]
\centering
        \subfigure[Movie.]{
		\begin{minipage}[b]{0.22\textwidth}
			\includegraphics[width=1\textwidth]{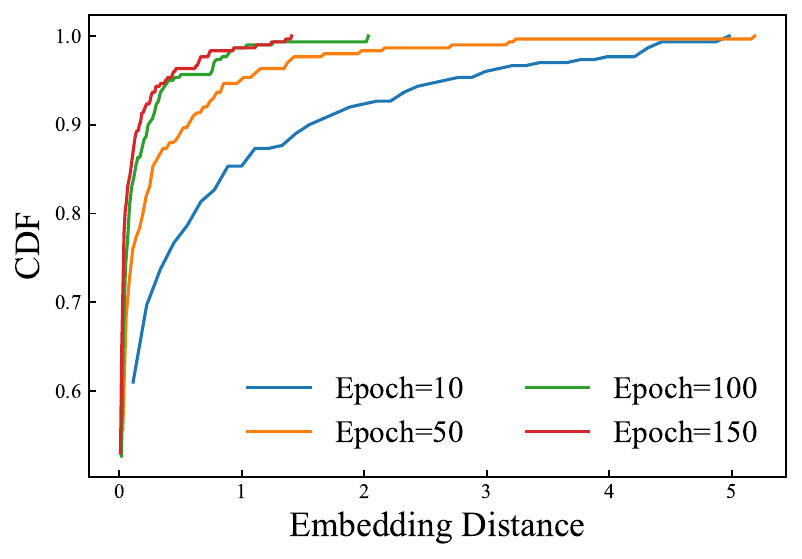} 
		
  \end{minipage}
		\label{movie_distance}
	}
        \subfigure[Stack.]{
            \begin{minipage}[b]{0.22\textwidth}
            \includegraphics[width=1\textwidth]{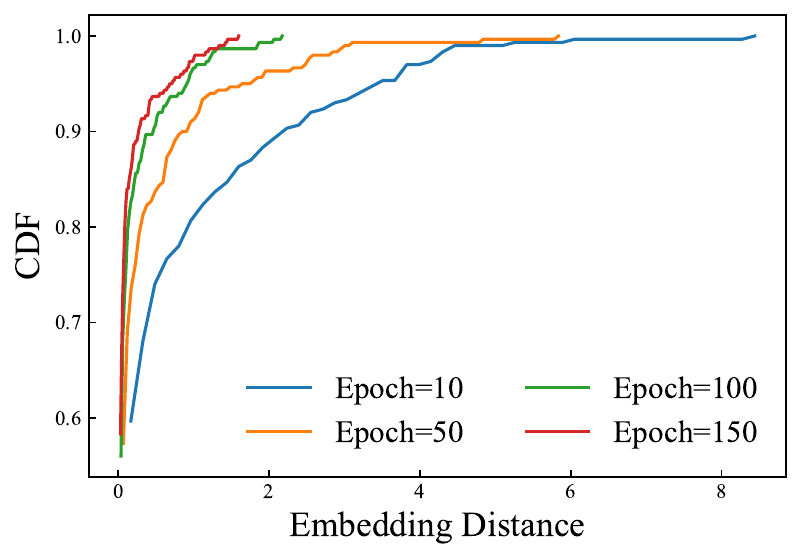}
            \end{minipage}
        \label{stack_distance}
        }
	\caption{\label{stale_comm}CDF of L2 distances between embeddings.}
\end{figure}

To further improve system efficiency by reducing network traffic, we propose adaptive stale embedding aggregation by exploiting the embedding similarity. This idea is motivated by an important observation that some vertices generate similar embeddings in different epochs. We collect all embeddings in some epochs when training DySAT models on the Movie and Stack datasets. The CDF of L2 distances between embeddings in a randomly selected epoch and those in the previous epoch is shown in Figure \ref{stale_comm}.
A smaller distance implies that embeddings have slight changes. We find that about 85\% embeddings in the 10-th epoch have distances less than 1 on the Movie dataset, while the corresponding percentage of the Stack dataset is 78\%. Moreover, we find that embedding distances become smaller as training proceeds. For example, over 95\% embeddings of the Movie dataset have distances less than 0.3 after 100 training epochs. 

The above observations motivate us to design a stale embedding aggregation mechanism. Specifically, we compare the current embeddings with the last transmitted ones, and transmit them only when their differences are sufficiently big. 
To calculate distances, it is only necessary to cache one copy of the embeddings for each vertex, specifically the last transmitted ones. Thus, the memory cost is affordable. Moreover, distances are calculated by the CPU, which has sufficient memory to accommodate these data. An alternative design is to compare the current embedding with the one in the preceding epoch, which would update the cached embeddings in every epoch. 
However, this design would accumulate embedding errors. Suppose consecutive embeddings have small distances, but accumulative distances between the first one and the final one would be big. This alternative design would never transmit embeddings, and may compromise training convergence. In contrast, our proposed method can well handle such accumulative embedding errors. 

Although stale aggregation can effectively reduce network traffic, it faces a critical challenge about how to decide the level of ``similarity'' for embedding reuse. If strict similarity requirements (e.g., an extreme case is that embeddings should be the exact same) are applied, many embeddings need to be transmitted over the network, and thus little traffic can be reduced. On the other hand, loose similarity requirements may decrease training accuracy. This trade-off is demonstrated by the experimental results shown in Table \ref{stale_acc}, where we let $D$ denote the maximum L2 embeddings distance in the current epoch and $\theta$ is a threshold of deciding whether embeddings can be reused. When we increase the value of $\theta$, the accuracy of all datasets decreases while more network traffic can be saved. In addition, we find that $\theta$ has different influences to different datasets. For example, when we set $\theta=0.3D$, about 85.5\% network traffic can be saved for Stack, with 0.076 drop in accuracy. To achieve a similar trade-off, we need to set $\theta=0.5D$ for the Movie dataset. The above fact demonstrates that a fixed value of $\theta$ cannot work well for all datasets, and we need to adaptively set it according to dataset characteristics.

Similar ideas of stale aggregation have been also adopted by \cite{wan2021pipegcn, miao2021het}. However, they define a maximum number of stale epochs that can be tolerated, instead of measuring the embedding similarity. This simple and straightforward method would decrease training accuracy. A recent work, \textsc{Sancus} \cite{peng2022sancus}, measures embedding similarity and defines a static handcrafted threshold to decide whether embeddings can be reused. 

\begin{table}[t]
\begin{center}
\resizebox{\linewidth}{!}{
\begin{tabular}{c|c|c|c|c|c|c|c}%
\hline  
\multirow{2}*{Dataset} & \multirow{2}*{Metrics} & \multicolumn{6}{c}{Static stale threshold} \\
\cline{3-8} & & $\theta=0$ & $\theta=0.1D$ & $\theta=0.3D$ & $\theta=0.5D$ & $\theta=0.7D$ & $\theta=0.9D$\\
\hline
\multirow{2}*{Amazon} & Accuracy & 0.690 & 0.688 & 0.680 & 0.677 & 0.658 & 0.644 \\
& Reduce comm. & - & 0.65\% & 54.14\% & 74.22\% & 87.87\% & 97.49\% \\
\hline
\multirow{2}*{Epinion} & Accuracy & 0.735 & 0.699 & 0.679 & 0.674 & 0.657 & 0.641 \\
& Reduce comm. & - & 6.27\% & 50.32\% & 75.15\% & 90.23\% & 98.38\% \\
\hline
\multirow{2}*{Movie} & Accuracy & 0.824 & 0.807 & 0.779 & 0.742 & 0.702 & 0.704 \\
& Reduce comm. & - & 44.64\% & 50.34\% & 80.80\% & 92.64\% & 97.67\% \\
\hline
\multirow{2}*{Stack} & Accuracy & 0.697 & 0.672 & 0.621 & 0.605 & 0.588 & 0.591 \\
& Reduce comm. & - & 51.36\% & 85.48\% & 87.56\% & 97.22\% & 98.98\% \\
\hline
\end{tabular}}
\end{center}
\caption{Test accuracy with different threshold $\theta$. 
}
\label{stale_acc}
\end{table}
 
We propose an adaptive stale embedding aggregation scheme to reduce communication cost while guaranteeing training convergence. Specifically, for each epoch $r$, we define a threshold $\theta_{r}$ of embedding similarity to determine whether embeddings could be reused. Its value is calculated by
\begin{align}
    &\theta_{r} = \frac{1}{1+\exp (norm(l_{r-1}))}D_{r}, \label{cal_theta}\\
    &norm(l_{r-1}) = \frac{l_{1}-l_{r-1}}{l_{1}},
\end{align}
where $l_{r-1}$ is the loss value of the epoch $r-1$ and $D_{r}$ is the maximum L2 distance among embeddings in the epoch $r$. Note that $l_{1}$ is the initial loss value when the training starts and the term $norm(l_{r-1}) = (l_{1}-l_{r-1})/l_{1}$ represents the normalized loss decrease in the epoch $r$. We use the scaled sigmoid function to adjust the threshold $\theta$. The rationale is as follows. In the early stages of training, the model is unstable, and we adopt a small $\theta$ to ensure most aggregated embeddings are fresh for quick training convergence. As the training progresses, the model tends to be stable, and we increase $\theta$ to reduce communication cost with trivial negative influence to training convergence.

\noindent\textbf{Discussion. }The design objective of the adaptive stale embedding aggregation technique is to strike a balance between network traffic and accuracy. Our experiments demonstrate that the communication costs are significantly reduced while there is a slight decrease in accuracy. Additionally, if there is a strict requirement for accuracy, DGC provides the option to disable the adaptive stale embedding aggregation, thereby maintaining the same accuracy as traditional distributed training systems.

\section{Implementation} \label{implementation}

\textbf{Dynamic Graphs and DGNN Models: }DGC is built on the top of PyTorch \cite{paszke2019pytorch} and PyTorch Geometric (PyG) \cite{fey2019fast}, which are widely used open-source frameworks for graph learning. In DGC, we represent dynamic graphs with \texttt{DynamicGraphSignal}, an iterator that divides the dynamic graph into multiple snapshots. Each snapshot is deployed with \texttt{data.Data}, defined in PyG. 
The DGNN models used in DGC are implemented with PyG and PyTorch APIs. Specifically, the GNN operations in the structure encoder (e.g., GCN and GAT) are implemented with PyG's APIs, such as \texttt{nn.conv.GCNConv} and \texttt{nn.conv.GATConv}. For RNN operations of the time encoder (e.g., GRU and LSTM), we implement them using \texttt{torch.nn.GRU} and \texttt{torch.nn.LSTM} provided by PyTorch.

\noindent\textbf{MLP Predictors: }we evaluate chunk workloads by two trained MLPs (\S\ref{chunk_assignment_sec}). Each MLP consists of an input layer, three hidden layers, and an output layer. We use 256 units in each hidden layer and a ReLU activation function after each layer. The output of the final layer is a single real number, which is the predicted execution time. Both MLPs have three input: (1) chunk information, i.e., number of vertices and number of edges; (2) feature information, i.e., vertex feature dimensions; (3) encoder information, i.e., layer dimensions; In this work, we focus on the homogeneous GPU setting. Thus, we do not add the GPU information to the MLP input. We randomly generate 50000 chunks offline from four dynamic graphs (Table \ref{Dataset}) and feed them into the structure and time encoder to measure their execution time as training labels. We adopt a mean absolute percentage error as the loss function and optimize MLPs with an Adam optimizer \cite{kingma2014adam}. Each MLP is trained for 100 epochs.

\noindent\textbf{Caching Module for Stale Aggregation: }we maintain KVStore servers and clients in GPU workers to cache vertex embeddings for remote aggregation. In our implementation, we deploy one KVStore server. Each GPU worker maintains a KVStore client and communicates with the KVStore server through \texttt{torch.distributed.rpc} APIs. Each GPU worker calls a \texttt{push()} API to send local embeddings to the KVStore server and updates the caching content. Meanwhile, the GPU worker can call a \texttt{pull()} API to aggregate remote embeddings from the KVStore server, and update contents cached in KVStore client.


\section{Evaluation} \label{evaluation}

\subsection{Experiment Setup}

\textbf{Environment settings \& Metrics. }We deploy DGC on a testbed consisting of eight NVIDIA Tesla V100 GPUs. We use Ubuntu 18.04 with Linux kernel version 5.4, NVIDIA driver 418.21, CUDA 10.1, and cuDNN 8.0.4. The versions of PyG and PyTorch are 2.0.4 and 1.11, respectively. In the overall performance comparison, we train the DGNN model for 100 epochs and measure average training time, including data loading, chunk fusion, remote communication, and GPU computation. We do not include the graph partitioning in the training time measurement since it can be executed offline before DGNN training. 

\noindent\textbf{Models \& Datasets. }We use the four datasets in Table. \ref{Dataset}. We divide these four datasets into snapshots with different window sizes, as being done by \cite{sankar2020dysat}. Specifically, Amazon contains graph data of 3650 days and we let the data of every 30 days be a snapshot. We set window sizes of Epinion, Movie, and Stack datasets as 1, 30, and 10, respectively. For all datasets, we use the in-degree and out-degree as the vertex features, similar to \cite{malik2021dynamic, chakaravarthy2021efficient}. We choose three representative DGNN models, whose details are as follows, to evaluate the performance of DGC.

\begin{itemize}
    \item \noindent\textbf{T-GCN} \cite{zhao2019t}: it utilizes three 2-layer GCN \cite{kipf2016semi} as the structure encoder, and a 1-layer GRU \cite{cho2014learning} model as the temporal encoder.
    
    \item \noindent\textbf{DySAT} \cite{sankar2020dysat}: it uses a 1-layer graph attention network (GAT) \cite{velivckovic2017graph} and a 1-layer scaled dot-product attention model \cite{vaswani2017attention} within each of its DGNN blocks.
    
    \item \noindent\textbf{MPNN-LSTM} \cite{malik2021dynamic}: it employs a 2-layer GCN and a 2-layer LSTM \cite{graves2012long} as the structure and time encoders, respectively. The structural embedding generation is similar with T-GCN. However, MPNN-LSTM uses a concatenation of outputs from each GCN layer as the input to the time encoder.
\end{itemize}

\noindent\textbf{Baselines. }We implement 3 baseline systems with different graph partitioning methods (e.g., PSS, PTS and PSS-TS), based on the state-of-the-art DGNN framework PyTorch Geometric Temporal (PyGT) \cite{rozemberczki2021pytorch}. They are referred to as PyGT-PSS, PyGT-PTS and PyGT-PSS-TS, respectively.

\subsection{Overall Performance Comparison}


\begin{figure}[t] 
\begin{center}
\includegraphics[width=0.46\textwidth]{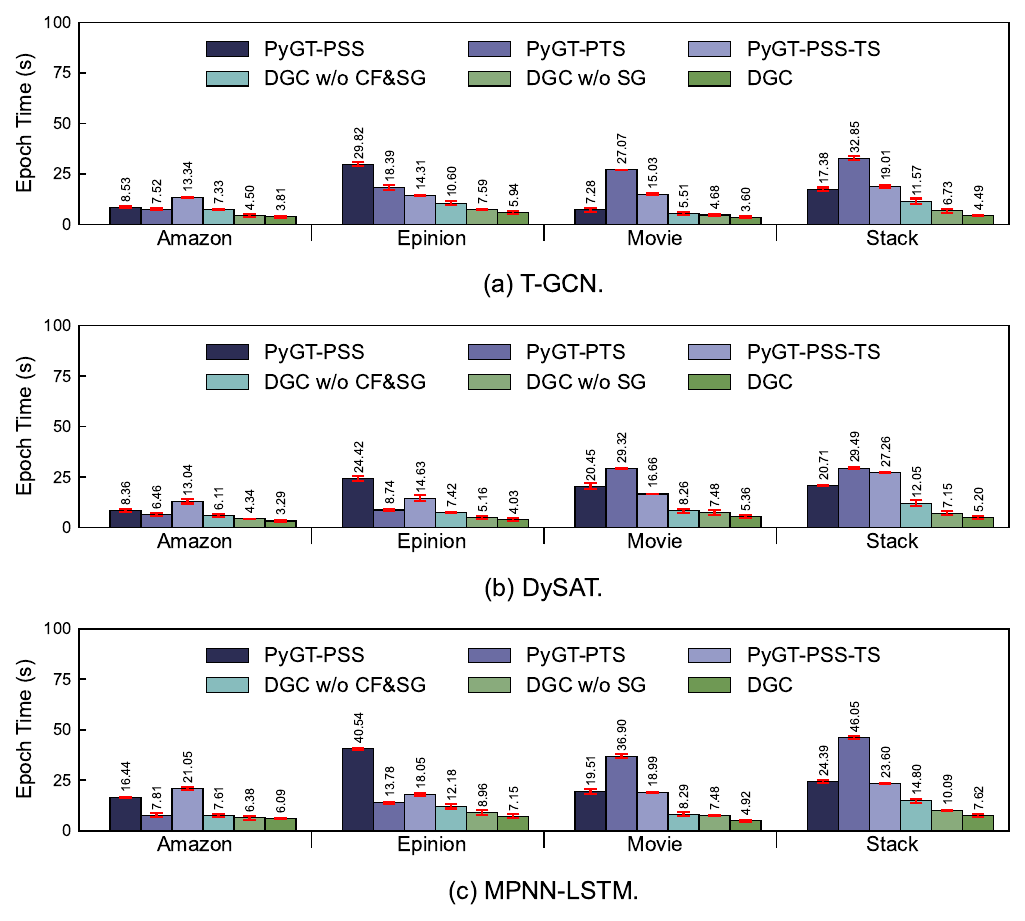}
\caption{\label{ablation_performance}Epoch time of different methods.}
\end{center}
\end{figure}

Figure \ref{ablation_performance} shows the overall speedup over PyGT when using different models. DGC outperforms all baselines by 1.25$\times$ - 7.52$\times$ (on average 3.95$\times$, 3.97$\times$, and 3.77$\times$ for T-GCN, DySAT, and MPNN-LSTM, respectively). Different baselines exhibit varying performance on these datasets.
Specifically, on dynamic graphs with fewer spatial dependencies (e.g., Amazon, with only 5.7M edges in total), PyGT-PTS, which breaks spatial dependency, performs better than PyGT-PSS and PyGT-PSS-TS. This is because fewer spatial dependencies lead to lower communication costs for PyGT-PTS. PyGT-PSS has the worst performance on the Epinion dataset. The reason is that the Epinion dataset has more snapshots, resulting in more temporal dependency. Therefore, PyGT-PSS incurs higher communication cost and longer epoch time, as it neglects the temporal features of dynamic graphs.
PyGT-PSS-TS avoids both spatial and temporal communication cost, but adds a shuffling cost to reassign embeddings to GPUs, which depends on the number of vertices. Therefore, for dynamic graphs with a large number of vertices (e.g., Amazon, with 103M vertices in total), it shows worse performance than the other methods. In contrast, DGC partitions dynamic graphs by chunks, considering both spatial and temporal features. It consistently outperforms other approaches and achieves the highest performance. In the following, we give details about results under different DGNN models.

\noindent\textbf{T-GCN. }Since T-GCN has two GCN layers and one GRU layer in each block, it involves more spatial communication than temporal communication.
DGC can exploit this characteristic to obtain more acceleration by reducing spatial communication cost. For instance, DGC achieves a 7.52$\times$ speedup compared to PyGT-PTS on the Movie dataset, where PyGT-PTS ignores the spatial features of dynamic graphs.

\noindent\textbf{DySAT. }In contrast to T-GCN, DySAT includes only one GAT layer and one temporal attention layer in the DGNN block. However, the unique self-attention mechanism in the temporal attention layer aggregates more temporal neighbors, compared to a GRU or LSTM layer. Specifically, in a GRU layer, each vertex only needs to aggregate the embeddings of its counterpart in the previous snapshot, while the counterparts in all snapshots should be aggregated by a temporal attention layer. The increased number of temporal neighbors results in higher temporal communication costs. Even though, DGC achieves a high speedup of 6.06$\times$ when training a DySAT on the Epinion dataset, compared to PyGT-PSS.

\noindent\textbf{MPNN-LSTM. }Because of two GCN layers and two LSTM layers in each block, MPNN-LSTM incurs higher communication costs than other models when training on the same dataset. DGC can reduce the epoch time by 7.5$\times$ compared to PyGT-PTS on the Movie dataset while achieving a speedup of 5.67$\times$ over PyGT-PSS on the Epinion dataset. However, the gap between PyGT-PSS-TS and DGC narrows when training MPNN-LSTM. The reason is that DGC's communication cost increases due to more layers adopted by MPNN-LSTM, but PyGT-PSS-TS's shuffling cost has almost no change.

\subsection{Ablation Study}

\subsubsection{Impact of PGC module}
\begin{figure}[t]
\centering
        \subfigure[Non-uniformity in spatial features.]{
		\begin{minipage}[b]{0.22\textwidth}
			\includegraphics[width=1\textwidth]{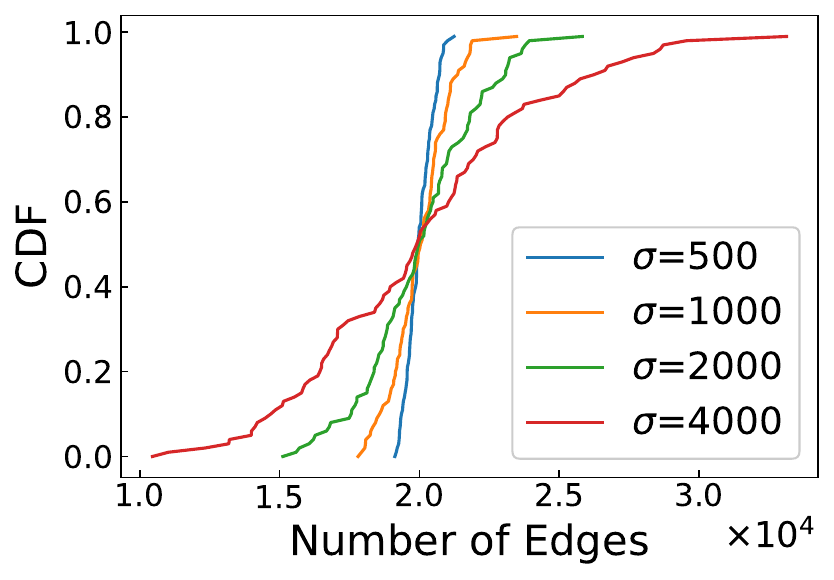}
            \end{minipage}
		\label{synthetic_spatial_cdf}
	}
        \subfigure[Non-uniformity in temporal features.]{
            \begin{minipage}[b]{0.22\textwidth}
            \includegraphics[width=1\textwidth]{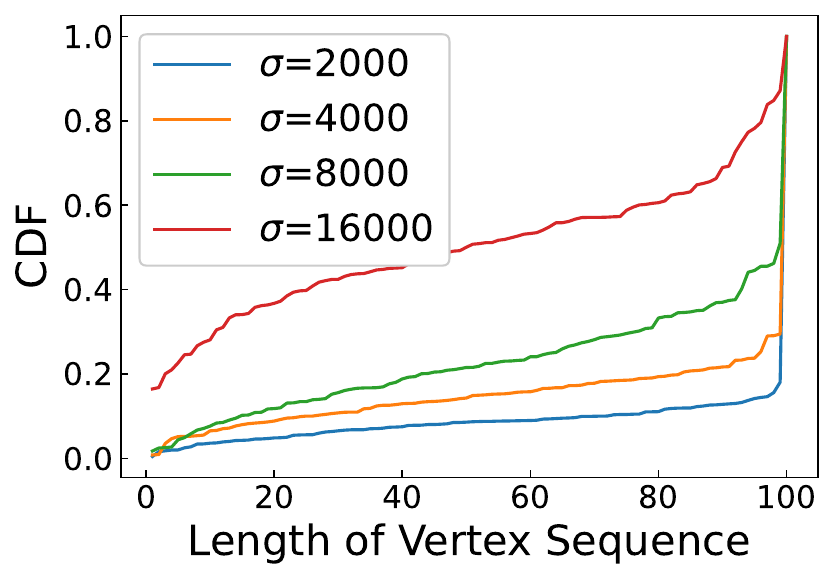}
            \end{minipage}
        \label{synthetic_temporal_cdf}
        }
	\caption{\label{synthetic_datasets}Synthetic datasets.}
\end{figure}
\begin{figure}[t]
\centering
        \subfigure[Impact of spatial non-uniformity.]{
		\begin{minipage}[b]{0.45\textwidth}
			\includegraphics[width=1\textwidth]{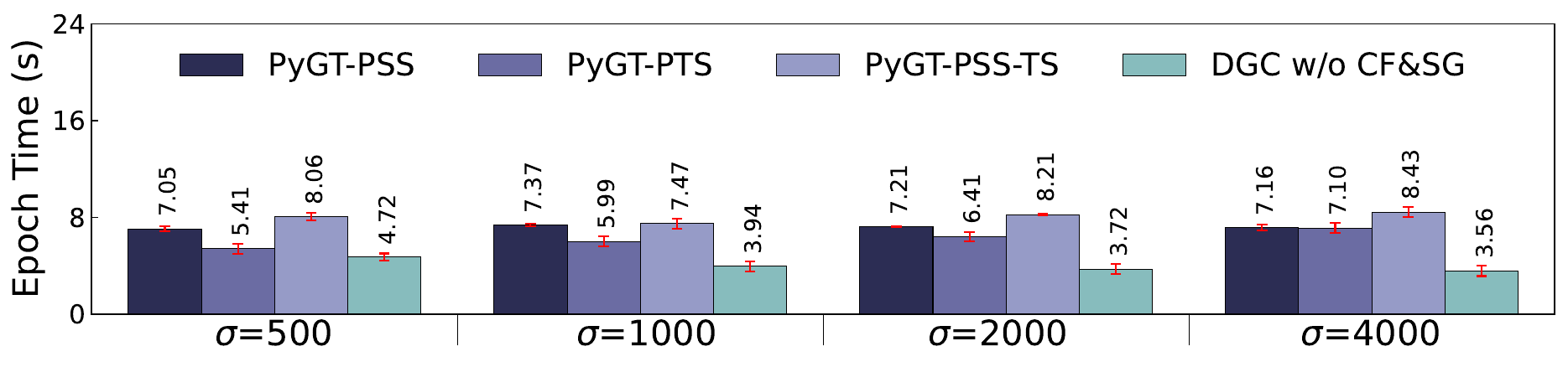}
            \end{minipage}
		\label{synthetic_spatial_performance}
	}
 \quad
        \subfigure[Impact of temporal non-uniformity.]{
            \begin{minipage}[b]{0.45\textwidth}
            \includegraphics[width=1\textwidth]{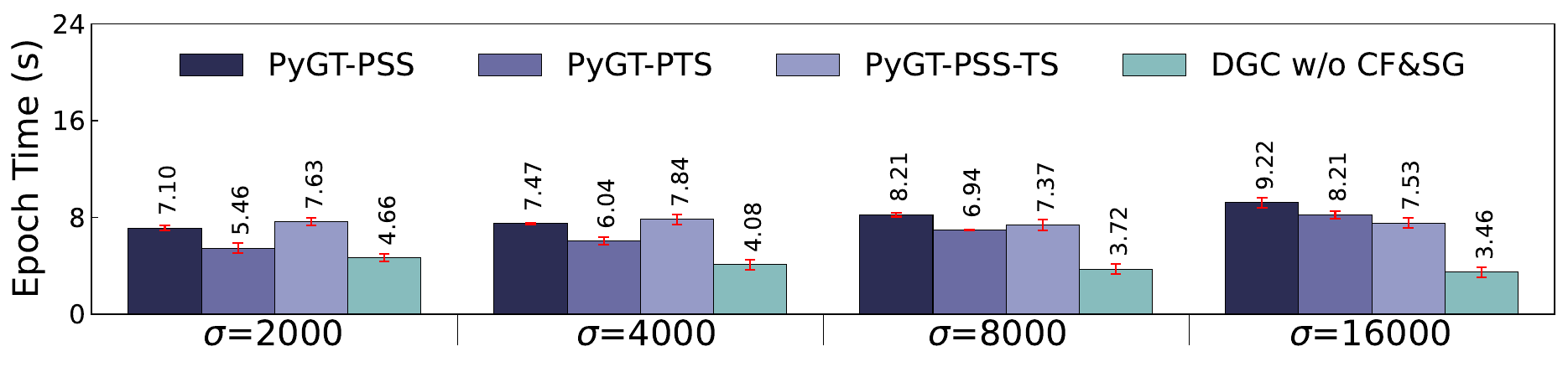}
            \end{minipage}
            \label{synthetic_temporal_performance}
        }
	\caption{\label{synthetic_performance}Impact of spatio-temporal non-uniformity.}
\end{figure}
As shown in Figure \ref{ablation_performance}, DGC with only PGC module, denoted by the bar of ``DGC w/o CF\&SG'', can still accelerate the training process by 1.03$\times$ to 4.92$\times$, compared to other methods. PyGT-PTS achieves similar performance with ``DGC w/o CF\&SG'' on Amazon dataset, because this dataset has very few spatial edges within each snapshot and the PTS can generate few cross-GPU traffic, leading to short epoch time. 

We further study how data non-uniformity affect the performance of PGC module on synthetic dynamic graphs. The synthetic graphs are generated by setting the total number of vertices, edges, and snapshots to 5M, 2M, and 100, respectively. In order to adjust non-uniformity levels of spatial features, we adjust the number of edges for each snapshot according to a normal distribution with a fixed mean value (i.e., 20K), and variable variances $\delta$, as shown in Figure \ref{synthetic_spatial_cdf}. In addition, we change the number of vertices in each snapshot to generate sequences of different lengths, so that we can study the influence of different non-uniformity levels in temporal features, as shown in Figure \ref{synthetic_temporal_cdf}.


As shown in Figure \ref{synthetic_performance}, in both cases, DGC with the PGC module always has the shortest epoch time and it decreases as the levels of spatial and temporal non-uniformity increase. This can be attributed to the following reason. The PGC module effectively reduces communication costs by aggregating vertices with important spatial features (i.e., those with more spatial dependencies but shorter vertex lengths) or significant temporal features (i.e., those with fewer spatial dependencies but longer sequence lengths). As the non-uniformity levels increases, spatial and temporal features become more obvious, which can be well handled by the PGC module. 
\subsubsection{Impact of chunk fusion module}
\begin{figure}[t]
\subfigure[\label{data_loading}Data loading cost.]{
\centering
\includegraphics[width=0.45\textwidth]{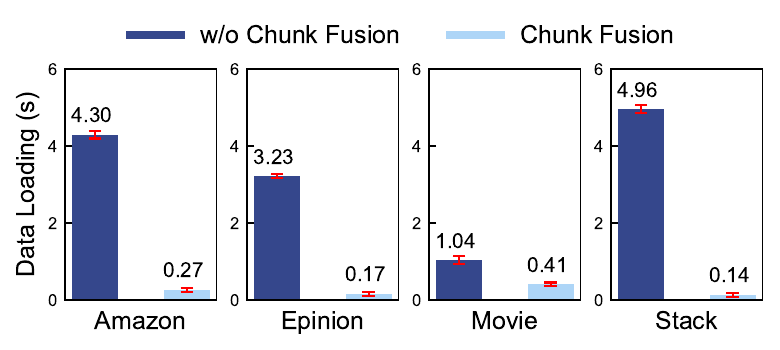}
}
\quad
\subfigure[\label{memory_util}GPU memory utilization.]{
\centering
\includegraphics[width=0.45\textwidth]{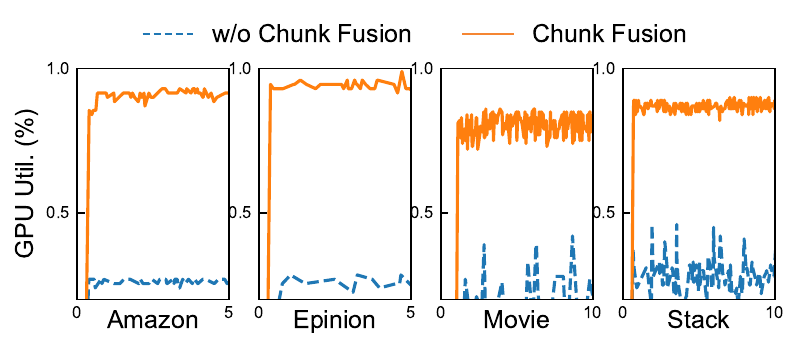}}
\caption{\label{impact_chunk_fusion}Chunk fusion performance.}
\centering
\end{figure}

As shown in Figure \ref{ablation_performance}, when chunk fusion (CF) is enabled, the epoch time can be further reduced by 1.39$\times$. In particular, the chunk fusion module is most effective on the Stack dataset. That is because this dataset is large and dense, resulting in large data loading cost, which can be effectively reduced by the chunk fusion module. 

To clearly show this benefit, we measure the data loading time per epoch when training MPNN-LSTM on four datasets with and without chunk fusion, and show results in Figure \ref{data_loading}. We can see that chunk fusion can significantly reduce data loading cost by 93.72\%, 94.74\%, 60.58\% and 97.18\% on Amazon, Epinion, Movie and Stack, respectively. Without chunk fusion, Stack dataset has the highest data loading cost because it has the largest dynamic graph size. Our chunk fusion searches chunks with the most data dependency to reduce redundant data loading. Moreover, we observe that the improvement on Movie dataset is smaller than that on other datasets. The reason is that vertex degrees of Movie dataset exhibits a significant power-law distribution, which implies that vertices with high degrees would be grouped to graph chunks with large sizes in chunk generation. Thus, generated chunks in Movie dataset have various sizes. The large chunks have few opportunities to be fused with others, leaving a small optimization space in chunk fusion. Thus, compared to other datasets, Movie gets limited improvement by chunk fusion.

Besides reducing data loading cost, chunk fusion lets multiple chunks be processed simultaneously to improve GPU utilization. Figure \ref{memory_util} shows the GPU utilization when training the MPNN-LSTM model with and without chunk fusion. As a result, chunk fusion significantly improves GPU utilization by 20\% to 95\%.

\subsubsection{Impact of adaptive stale aggregation module}\label{impact_SG}
\begin{table*}[h]
\begin{center}
\label{exp_stale}
\begin{tabular}{c|c|c|cc|cc|cc|cc}%
\hline  
\multirow{3}*{Dataset} & \multirow{3}*{Model} & \multicolumn{1}{c|}{w/o stale aggre.} & \multicolumn{8}{c}{with stale aggre.} \\
\cline{3-11} & & \multirow{2}*{Acc} & \multicolumn{2}{c|}{$\theta=0.3D$} & \multicolumn{2}{c|}{$\theta=0.5D$} & \multicolumn{2}{c|}{$\theta=0.7D$} & \multicolumn{2}{c}{Adaptive threshold}\\
\cline{4-11} & & & \multicolumn{1}{c}{Acc} & \multicolumn{1}{c|}{Comm.} & Acc & Comm. & Acc & Comm. & Acc & Comm. \\
\hline

\multirow{3}*{Amazon} & T-GCN & 
 \textbf{0.667} &
 0.653 & 40.39\% & 0.652 & 48.74\% & 0.639  & 85.97\% &
 0.668 & 79.32\%
\\ 
& DySAT & 
 \textbf{0.685} &
 0.687 & 46.40\% & 0.657 & 55.63\% & 0.636 & 77.05\% &
 0.656  & 73.03\% 
\\
& MPNN-LSTM & 
 \textbf{0.674} &
 0.653 & 53.03\% & 0.648  & 37.96\% & 0.631 & 86.47\% &
 0.654  & 79.31\%
\\
\hline

\multirow{3}*{Epinion} & T-GCN & 
 \textbf{0.732} &
 0.703 & 46.4\% & 0.691 & 50.17\% & 0.651 & 97.31\% &
 0.701 & 81.44\%
\\ 
& DySAT & 
 \textbf{0.738} &
 0.674 & 63.49\% & 0.660 & 69.71\% & 0.648 & 93.34\% &
 0.702 & 79.17\%
\\
& MPNN-LSTM & 
 \textbf{0.661} &
 0.630 & 68.98\% & 0.635 & 69.29\% & 0.607 & 97.47\% &
 0.630 & 94.11\%
\\
\hline

\multirow{3}*{Movie} & T-GCN & 
 \textbf{0.839} &
 0.828 & 56.39\% & 0.825 & 57.84\% & 0.774 & 96.07\% &
 0.830 & 77.68\%
\\ 
& DySAT & 
 \textbf{0.829} &
 0.819 & 62.81\% & 0.812 & 70.98\% & 0.781 & 93.36\% &
 0.819 & 78.87\%
\\
& MPNN-LSTM & 
 \textbf{0.727} &
 0.723 & 31.94\% & 0.721 & 32.96\% & 0.623 & 83.10\% &
 0.723 & 53.08\%
\\
\hline
\multirow{3}*{Stack} & T-GCN & 
 \textbf{0.698} &
 0.702 & 45.56\% & 0.696 & 55.36\% & 0.599 & 97.91\% &
 0.694 & 86.23\%
\\ 
& DySAT & 
 \textbf{0.703} &
 0.700 & 68.45\% & 0.691 & 79.68\% & 0.602 & 95.78\% &
 0.699 & 89.02\%
\\
& MPNN-LSTM & 
 \textbf{0.654} &
 0.654 & 67.95\% & 0.651 & 68.14\% & 0.553 & 97.70\% &
 0.644 & 91.76\%
\\
\hline
\end{tabular}
\end{center}
\caption{Impact of adaptive stale embedding aggregation. 
} 
\label{impact_stale}
\end{table*}

We finally enable the adaptive stale aggregation module. As shown in Figure \ref{ablation_performance}, we find that the training time shows even more improvement, by 1.32$\times$ faster. This module dramatically cuts down the volume of communication, thereby expediting the training.

To better understand the benefits of adaptive stale aggregation module, we further conduct experimental comparison with static stale thresholds. We have three different settings for static thresholds: $\theta=0.3D$, $0.5D$, and $0.7D$. The test accuracy and reduced communication cost when training T-GCN, DySAT, and MPNN-LSTM on four datasets are shown in Table \ref{impact_stale}. In addition, we also show the results of a DGC variant without stale aggregation, so that we can clearly show the effectiveness of our stale aggregation.

The stale embedding aggregation can reduce communication cost by 32.96\% to 97.70\% compared to the DGC without stale embedding aggregation. 
Recall that the design objective of the adaptive stale embedding aggregation is to strike a balance between network traffic and accuracy. According to Table \ref{impact_stale}, the average accuracy drop of our proposed method is 1.56\%, but it saves 80.26\% cross-GPU traffic, which can significantly accelerate the training speed. We believe this design is attractive to users who care about the time-to-accuracy metric. Moreover, since adaptive stale aggregation is a pluggable module, we can disable it for users who strongly care about accuracy.
When training a DySAT model on the Stack dataset, the stale embedding aggregation with a static threshold of $0.5D$ can decrease communication costs by 79.68\%. However, the benefit of traffic reduction is only 32.96\% when training the MPNN-LSTM model. Furthermore, a static stale threshold is inadequate in maintaining training convergence. In the Epinion dataset, with a static threshold ($\theta=0.5D$), the communication cost is reduced by 69.71\%. However, the test accuracy decreases from 0.738 to 0.66. 


\subsection{Chunk workload prediction. }
\begin{figure*}[t] 
\begin{center}
\includegraphics[width=0.95\textwidth]{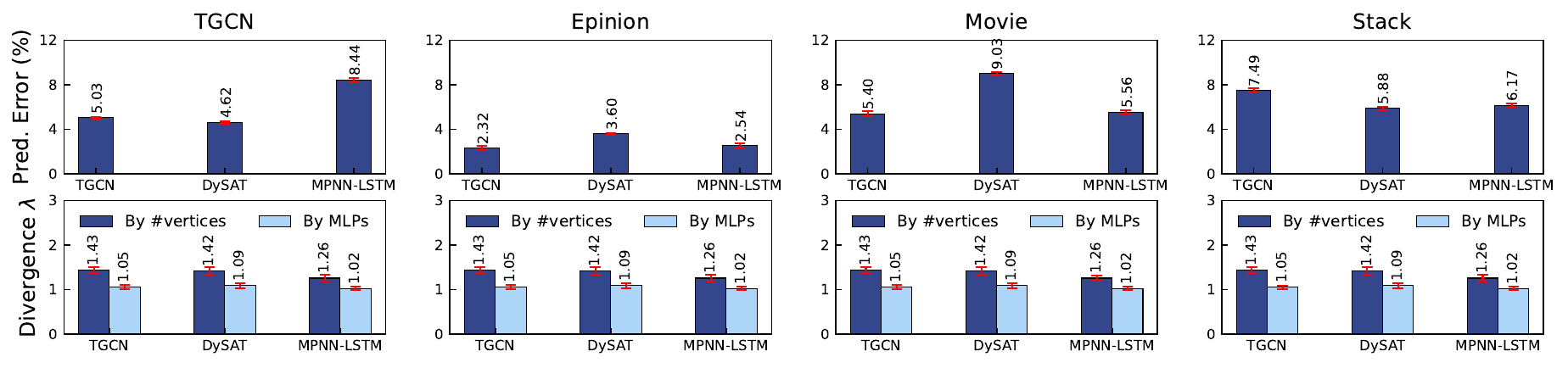}
\caption{\label{impact_prediction}Chunk workload prediction error and workload divergence.}
\end{center}
\end{figure*}
Since chunk workload estimation greatly affects the workload balance in chunk assignment, we evaluate the accuracy of our proposed MLPs that predict the execution time of chunks (\S\ref{chunk_assignment_sec}). We randomly choose graph chunks from four datasets and compare their measured execution time and predicted one by MLPs, which are denoted by $\text{measured}_{a}$ and $\text{predicted}_{a}$, respectively. We define the prediction error as: 
\begin{align}
    error = \frac{1}{n}\sum\limits_{a=1}^{n}\frac{|\text{predicted}_{a} - \text{measured}_{a}|}{\text{measured}_{a}}.
\end{align}
We set $n=1000$ in our experiments. As shown in Figure \ref{impact_prediction}, the prediction error is less than 10\%, which demonstrates the proposed MLPs have sufficient accuracy to estimate chunk workloads.

We further study the impact of workload prediction on chunk assignment. As shown in Figure \ref{impact_prediction}, we report the workload divergence (defined in \S\ref{motivation_1}) of two workload prediction methods. The first one is the baseline method, which estimates the workloads of graph chunks by counting the number of vertices \cite{stanton2012streaming,lin2020pagraph}. The second method is to evaluate workloads by MLPs, which is adopted by DGC. The results show that trained MLPs can achieve better workload balance. Specifically, the average workload divergence is about 1.23 when using MLPs, while the divergence increases to 1.67 when we use the number of vertices as chunk workloads.

\subsection{DGC Overhead}
\begin{figure}[t] 
\begin{center}
\includegraphics[width=0.45\textwidth]{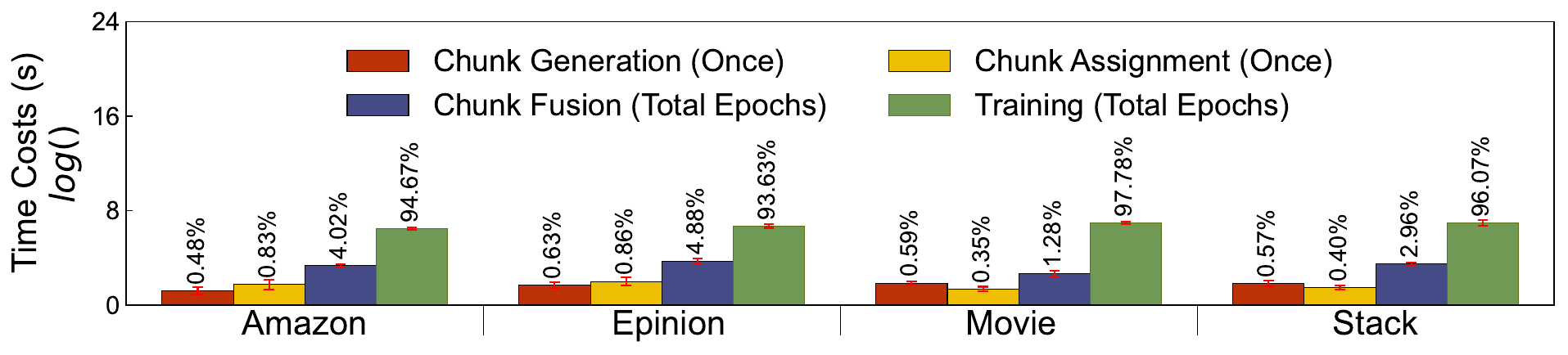}
\caption{\label{addition_costs}Extra overhead introduced by DGC. The numbers above bars are the percentages of total training time.}
\end{center}
\end{figure}

To analyze the extra overhead introduced by DGC, we measure graph partitioning overhead (including chunk generation and chunk assignment) and chunk fusion overhead, as depicted in Figure \ref{addition_costs}. Note that graph partitioning is only invoked once per training job. Due to the substantial gap between different operations, we use a logarithmic function to normalize the overhead. The results indicate that DGC introduces only about 4\% overhead to the total training time. Additionally, we find that the Amazon dataset has the lowest chunk generation overhead since it possesses the fewest edges, allowing label propagation to converge swiftly. However, Amazon dataset has a high chunk assignment cost. That is because Amazon dataset has a large number of vertices, resulting in lots of generated chunks and high chunk assignment cost. Although the total chunk fusion overhead exceeds the graph partitioning overhead, it is still significantly less than the total training time. Consequently, the additional overhead introduced by DGC does not significantly detract from the overall performance.

\subsection{Convergence Evaluation}
\begin{figure*}[t] 
\begin{center}
\includegraphics[width=0.98\textwidth]{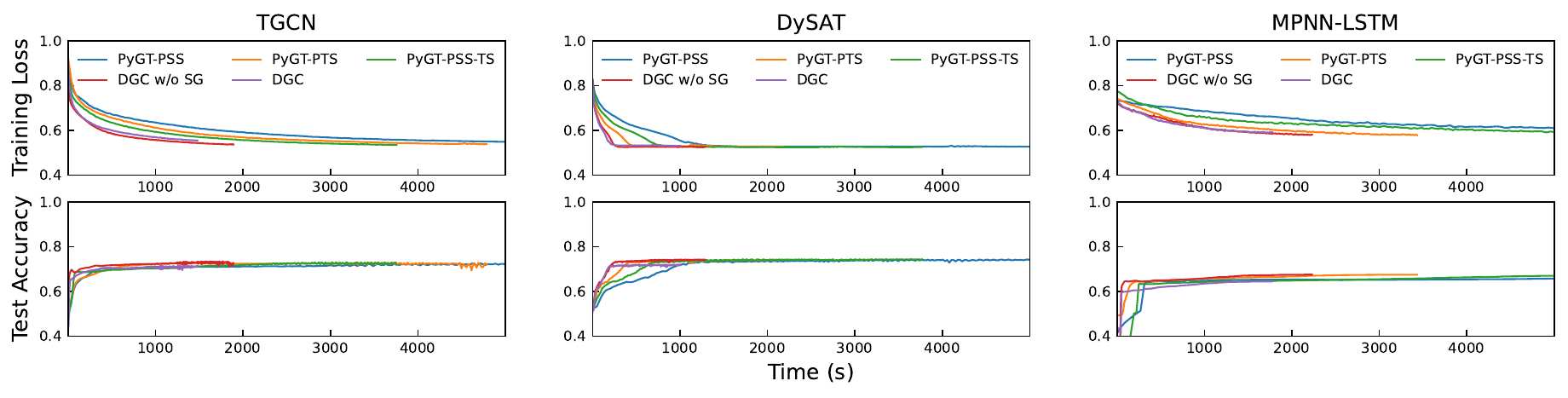}
\caption{\label{training_convergence}Training loss and test accuracy under different methods for three models on the Epinion dataset.}
\end{center}
\end{figure*}

Finally, we study the training convergence of DGC. Figure \ref{training_convergence} shows the curves of training loss and test accuracy when we train different models on Epinion dataset. We have similar observations on other datasets and thus omit their figures due to space limit, but corresponding accuracy is included in Table \ref{impact_stale}. We can see that all methods, except DGC with stale aggregation, can finally achieve similar accuracy and loss. That is because all methods adopt full-batch training, without changing training algorithms and related hyper-parameters. The results validate that DGC can guarantee training convergence. In addition, thanks to PGC and the chunk fusion modules, DGC can converge at a faster speed, compared to others. When adaptive stale aggregation module is enabled, DGC can further accelerate the convergence, with minor accuracy degradation.

\section{Related Work}  \label{related_work}


\textbf{Optimization for GNNs. }
Numerous studies have focused on scaling Graph Neural Networks (GNNs) for large graph training, which can be divided into two main categories. The first category focuses on algorithmic approaches, where existing works have explored various techniques to scale GNN models. These techniques include sampling methods \cite{chiang2019cluster, zeng2019graphsaint, zou2019layer, you2020l2, chen2021fedgraph}, quantization \cite{bahri2021binary, wang2021bi, WangQGTC, liu2021exact}, simplification \cite{wu2019simplifying, he2020lightgcn}, and distillation \cite{deng2021graph, yan2020tinygnn, zhang2020reliable}. The second category emphasizes distributed training, where GNN training is conducted using multiple CPUs or GPUs to manage large graphs \cite{ma2019neugraph, jia2020improving, liu2021glint, cai2021dgcl, md2021distgnn, gandhi2021p3, wang2021gnnadvisor, zhu2019aligraph, lin2020pagraph, liu2021bgl, wang2021flexgraph, thorpe2021dorylus, yang2022gnnlab}.
Several works have built upon general runtime frameworks, such as DGL \cite{wang2019dgl}, PyG \cite{fey2019fast}, and AGL \cite{zhang13agl}, to propose various optimizations. AliGraph \cite{yang2019aligraph} and  AGL \cite{zhang13agl} only support distributed GNN training on CPUs, while others \cite{ma2019neugraph, lin2020pagraph, wang2022neutronstar} support GNN training on GPUs. 
DistDGL \cite{zheng2020distdgl} optimizes graph data access by supporting a distributed in-memory key-value store. 
DGCL \cite{cai2021dgcl} improves distributed GNN training efficiency with an efficient communication library and NVLink. Roc \cite{jia2020improving} minimizes data swapping between GPU memory and host DRAM by using dynamic programming. 
$P^{3}$ \cite{gandhi2021p3} jointly combines intra-layer model parallelism and data parallelism to avoid communication costs for data-intensive node features among GPUs. Dorylus \cite{thorpe2021dorylus} deploys GNNs with serverless computing and increases training scalability at a low cost.

\noindent\textbf{Optimization for DGNNs. }Two general frameworks, PyGT \cite{rozemberczki2021pytorch} and TGL \cite{zhou2022tgl}, have been proposed to implement a variety of DGNN models. CacheG \cite{li2021cache} improves DGNN training performance by introducing intermediate result caching.
Cambricon-G \cite{song2021cambricon} combines a dedicated architecture, featuring a cuboid engine and hybrid on-chip memory, to decrease energy consumption and on-chip memory access for dynamic GNNs. TGOpt \cite{wang2023tgopt} specifically targets attention-based DGNNs and introduces a range of optimizations, such as deduplication, memorization, and precomputation, to minimize redundant computation during DGNN inference. PiPAD \cite{wang2023pipad} aims to enhance training efficiency and reduce data transfer overhead in the traditional ``one-graph-at-a-time'' DGNN training pattern. However, these works only support optimization for single-GPU DGNN training. In contrast, DGC focuses on efficient distributed DGNN training for handling large dynamic graphs.

\noindent\textbf{Graph Partitioning. }Partitioning graphs across multiple GPUs is essential to minimize cross-GPU traffic during distributed graph training. The graph partitioning problem has been extensively studied in distributed GNN training. For example, DistDGL \cite{zheng2020distdgl}, AliGraph \cite{zhu2019aligraph}, and DistGNN \cite{md2021distgnn} adopts the Metis partitioning algorithm \cite{karypis1997metis} to optimize cross-GPU communication costs. NeuGraph \cite{ma2019neugraph} adopts the Kernighan-Lin algorithm and Roc \cite{jia2020improving} uses a linear-regression based algorithm to partition graphs. $P^{3}$ \cite{gandhi2021p3} independently partitions the input graph and features to avoid communicating huge features over the network.
However, these graph partitioning methods do not apply to dynamic graph partitioning since they are designed for unraveling spatial dependency. Recently, several works have been proposed for dynamic graph partitioning in distributed DGNN training.
DynaGraph \cite{guan2022dynagraph} partitions the dynamic graph by temporal sequences, which effectively eliminates temporal embedding transmissions. Chakaravarthy et al. \cite{chakaravarthy2021efficient} propose a joint partitioning method that applies PSS to assign snapshots to GPUs to execute structure encoders, and then shuffling to PTS for running time encoders.
However, existing partitioning methods may not be suitable for various datasets, as they do not account for spatio-temporal non-uniformity in dynamic graphs. In contrast, DGC introduces a partitioning method based on graph chunks that takes full advantage of spatio-temporal non-uniformity in dynamic graphs. This approach leads to better workload balancing and reduced communication costs, significantly improving the DGNN training efficiency.

\section{Conclusion}  \label{conclusion}

This paper introduces DGC, a distributed training framework designed to optimize DGNN training efficiency. By incorporating a novel dynamic graph partitioning method (PGC) and run-time optimizations, DGC effectively tackles the challenges of high communication costs and low GPU utilization in distributed DGNN training. Experimental results demonstrate that DGC achieves a 1.25$\times$-7.52$\times$ speedup compared to state-of-the-art DGNN training frameworks.

\bibliographystyle{ACM-Reference-Format}
\bibliography{ref}

\end{document}